\documentclass{elsarticle}

\usepackage{etex}
\usepackage[english]{babel} 
\usepackage[latin1]{inputenc} 
\usepackage{latexsym}
\usepackage{epsfig} 
\usepackage{listings}
\usepackage{url,xspace,courier} 
\usepackage[draft]{fixme}
\usepackage{color} 
\usepackage{multirow}
\usepackage{latexsym,amssymb}
\usepackage{fancybox,epic}
\usepackage[noend]{algpseudocode}
\usepackage{amsmath}
\usepackage{graphics}
\usepackage{graphicx}
\usepackage{array}
\usepackage{mathpartir}

\usepackage{subcaption}

\usepackage{wrapfig}
\usepackage{epsfig}

\usepackage{booktabs}
\usepackage[all]{xy}

\usepackage{amssymb}
\usepackage{latexsym}
\usepackage{fancybox,epic}
\usepackage{algorithm}
\usepackage{amsmath}
\usepackage{colortbl}
\usepackage{enumitem}
\usepackage{pgf}
\usepackage{tikz}
\usepackage{amsmath}
\usepackage{cancel}

\usepackage{marvosym}

\usepackage{listings}
\lstdefinestyle{PrologStyle} {
  language=Prolog,
  framerule=0pt,
  basicstyle=\footnotesize\ttfamily,
  commentstyle=\color{black},
  keywordstyle=\footnotesize\ttfamily\bfseries,
  stringstyle=\footnotesize\ttfamily,
  showstringspaces = false,
  escapechar=@,
  mathescape=true,
  extendedchars=true,%
  lineskip=0ex,%
  framerule=0pt,
  numbersep=2pt,
  numberstyle=\tiny,
  deletekeywords={length,ef}, 
  morekeywords={set_field,get_field,return,async,get,new_object},
}

\newcommand{\key}[1]{\mbox{\ttfamily\bfseries #1}}

\newcommand{\buffer}{\mathit{a}}
\newcommand{\queue}[0]{\C{Q}}
\newcommand{\microstepstar}[1]{\stackrel{#1}{\leadsto^*}}
\newcommand{\macrostep}[1]{\stackrel{#1}{\longmapsto}}
\newcommand{\newheap}{\ensuremath{newheap}}
\newcommand{\C}[1]{{\cal #1}}
\newcommand{\rrderiv}{\stackrel{\objid\cdot\tkid}{\leadsto}}
\newcommand{\sepo}{\cdot}
\newcommand{\many}[1]{\overline{#1}}

\newcommand{\heap}{h}

\newtheorem{theorem}{Theorem}[section]

\newtheorem{example}[theorem]{Example}
\newtheorem{definition}[theorem]{Definition}
\newtheorem{lemma}[theorem]{Lemma}

\newenvironment{proof}{\noindent{\em Proof}.\\
}{\hfill$\Box$\vskip\partopsep \vskip\topsep}

\newcommand{\objid}[0]{\ensuremath{\mathit{o}}\xspace}

\newcommand{\tkid}[0]{\ensuremath{\mathit{tk}}\xspace}

\newcommand{\lst}[1]{\lstinline!#1!}

\newcommand{\selectObject}{\ensuremath{\mathit{selectAct}}\xspace}
\newcommand{\selectTask}{\ensuremath{\mathit{selectTask}}\xspace}

\newcommand{\Get}[1]{\ensuremath{\mbox{\bf \lstinline!get!}}\xspace}
\newcommand{\newb}{\ensuremath{\mbox{\lstinline!new!}}\xspace}

\newcommand{\iftethen}{{\ensuremath{\mbox{\lstinline!then!}}\xspace}}
\newcommand{\ifteelse}{{\ensuremath{\mbox{\lstinline!else!}}\xspace}}
\newcommand{\while}{\mbox{\lstinline!while!}\xspace}
\newcommand{\whilebody}{\mbox{\lstinline!do!}\xspace}

\newcommand{\await}[1]{\mbox{\bf \lstinline!await!}~#1?\xspace}
\newcommand{\returnval}[1]{\mbox{\lstinline!return #1!}\xspace}
\newcommand{\body}[1]{\ensuremath{\mathit{body}(#1)}}

\newcommand{\ex}{E}
\newcommand{\extend}[1]{S}

\newcommand{\syco}{SYCO\xspace}

\newcommand{\secbeg}{}

\newcommand{\subsecbeg}{}

\newcommand{\taskId}{{\sf tk}}
\newcommand{\task}{\textsc{tk}}
\newcommand{\tv}{{\it l}}
\newcommand{\state}[0]{S}
\newcommand{\futvar}[1]{{\tt #1_{\tt f}}}

\newcommand{\Fields}{{\it fields}}

\begin{document}

\begin{frontmatter}
  
\title{Actor-Based Model Checking for SDN Networks}

\cortext[cor1]{Corresponding author: Miguel Isabel, Department of Sistemas
Inform\'aticos y Computaci\'on , C/ Profesor Jos\'e Garc\'ia Santesmases, s/n
Complutense University of Madrid, E-28040 - Madrid (Spain). Phone/Fax +34 91
3947641 / +34 91 3947529.}

\author[1]{Elvira Albert}\ead{elvira@sip.ucm.es}
\author[1]{Miguel G\'omez-Zamalloa}\ead{mzamalloa@ucm.es}
\author[1]{Miguel Isabel \corref{cor1}}\ead{miguelis@ucm.es}
\author[1]{ Albert Rubio}\ead{albert@cs.upc.edu}
\author[2]{ Matteo Sammartino}\ead{m.sammartino@ucl.ac.uk}
\author[2]{ Alexandra Silva}\ead{alexandra.silva@ucl.ac.uk}

\address[1]{Complutense University of Madrid,  Spain}
\address[2]{University College London,  UK}

\setcounter{page}{1}




\begin{abstract}
  Software-Defined Networking (SDN) is a networking paradigm
  that has become increasingly popular in the last decade. The 
  unprecedented control over the global behavior of the network it provides opens a range of 
new opportunities for formal methods and  much work has
  appeared in the last few years on providing bridges between SDN and
  verification.  This article advances this research line and provides
  a link between SDN and traditional work on formal methods for
  verification of concurrent and distributed software---actor-based modelling. We
  show how SDN programs can be seamlessly modelled using
  \emph{actors}, and thus existing advanced model checking techniques
  developed for actors can be directly applied to verify a range of
  properties of SDN networks, including consistency of flow tables,
  violation of safety policies, and forwarding loops. Our model
  checker for SDN networks is available through an online web
  interface, that also provides the SDN actor-models for a number
  of well-known SDN benchmarks.
\end{abstract}


\end{frontmatter}



\section{Introduction}

SDN is a relatively recent networking paradigm which is now widely used in
industry, with many companies---such as Google and Facebook---using SDN to control their backbone networks and data-centers. The core principle in SDN is 
the separation of control and data planes---there is a centralized {\em controller} 
which operates a collection of distributed interconnected switches. The controller can 
 dynamically update switches'  policies depending on the observed flow of packets, which is a simple but powerful way to react to unexpected events in the network. 
Network verification has gained an extra boost since SDN was introduced, as in this new paradigm the amount of detailed information available about network events is rich enough and can be centrally gathered to check for properties, both statically and dynamically, of the network behavior. Moreover, the controller itself is a program which can be analyzed and verified before deployment.

The distributed and concurrent nature of network behavior makes programming and  verification tasks challenging. Some of the bugs that can be found in existing (programmable) networks are reminiscent of faults that have appeared in distributed and concurrent systems, and which have  inspired much research in the verification and formal methods communities. 
With this observation as a starting point, this article provides a new bridge between SDN and a strand of formal
methods---actor-based modelling \cite{actors}--- which 
 was originally developed to analyze concurrent systems. Actors, entities  equipped with a
private memory, form the
basic unit of computation in such framework and can interact with each other through
\emph{asynchronous} messages. This setup enables reasoning about local
properties of the system without knowledge of the whole program, which gives rise to
more compositional and thus scalable methods. Actors provide the
foundations for the concurrency model of languages used in industry,
e.g., \emph{Erlang} and \emph{Scala}, and libraries used in mainstream
languages, e.g., \emph{Akka}.

\subsection{Summary of contributions} 
This article makes five main contributions:
\begin{enumerate}
\item SDN-semantics: A formalization of the semantics of SDN
  networks which allows us to define the transitions that occur in the
  network and formalize the concept of execution trace
  needed to prove soundness of our modelling.

\item SDN-Actors: An encoding of all basic components of an SDN
  network (switches, hosts, controller) into the actor-based language ABS
  \cite{DBLP:conf/fmco/JohnsenHSSS10} and a soundness proof of our
  encoding using the semantics of SDN networks in point 1.

\item Barriers: One of the most challenging aspects to encode are the
  OpenFlow \emph{barrier} messages, special instructions that the
  controller can use to force switches to execute all their queued
  tasks. We provide an implementation of barriers using conditional
  synchronization and a soundness result.

\item Model checker: A model checker for our SDN models built on top
  of the SYCO tool \cite{DBLP:conf/cc/AlbertGI16} that incorporates
  several dynamic partial-order reduction (DPOR) algorithms.

\item Case studies: Several case studies of SDN and properties to
  illustrate the versatility and potential of the approach. We were
  able to find bugs related to programming errors in the controller,
  forwarding loops, and violation of safety policies, and scale to
  larger networks than related techniques.
\end{enumerate}
This article extends and improves the conference paper that appeared
in the FM'18 proceedings \cite{AlbertGRSS18} as follows. On the
theoretical side, we have formalized the semantics of SDN networks and
used it to prove soundness of the basic encoding of SDN-Actors,
ensuring thus the correctness of our models. On the practical side, we
have carried out a new experimental evaluation using the
Constrained DPOR algorithm \cite{DBLP:conf/cav/AlbertGIR18}. This DPOR
algorithm can take advantage of independence conditions that we have
defined specifically for the SDN domain and that allow us to treat
larger networks than by using related techniques and than in our FM'18
paper. We have also extended the SYCO tool with a mechanism to detect
the violation of the property under check that stops the exploration,
while before SYCO was restricted to full exploration.

\subsection{Organization of the article}

Section~\ref{sec:overview} gives an
intuition of the main ideas in the article by means of a simple
example. In Section~\ref{sec:semant-sdn-netw} we present the semantics
of SDN programs and of actor systems, in two parts. First,
Section~\ref{sec:sdn-networks} introduces a semantics for SDN networks
that describes the communication patterns in this kind of networks and
that allows us to formalize the notion of execution trace in the SDN
network. Next, we recall the semantics of actor systems from
\cite{DBLP:conf/fmco/JohnsenHSSS10} which will constitute the
semantics of our models. Section~\ref{sec:sdn-actors:-an} 
introduces the concept of \emph{SDN-Actor} by providing the encoding of
all components in an SDN network as actors. We formally prove the
soundness of the encoding by relying on the semantics introduced in
Section~\ref{sec:sdn-networks}. Section~\ref{sec:barriers} extends
both the SDN semantics and our models to handle barriers and formalizes
the soundness of this extension. Section~\ref{sec:dpor-based-model}
describes our DPOR-based model checker which instantiates an
off-the-shelf model checker for actor systems with tailored
independence conditions to efficiently verify SDN-Actor models. Section~\ref{sec:verif-sdn-prop}  describes the 
experimental evaluation of the tool. Related work and conclusions appear in Section~\ref{sec:concl-relat-work}.

\section{Overview} \label{sec:overview}
\secbeg

This section contains an overview of the technical contributions via
an extended example, which we also use to introduces basic concepts and notations.

\secbeg
\secbeg
\secbeg \vspace{-.1cm}
\subsection{Concurrency errors in SDN networks}
\secbeg

SDN is a networking architecture where a central software \emph{controller} 
can dynamically change how network switches forward packets by monitoring the traffic.
Switches can be connected to hosts and to other switches via bidirectional channels 
that may reorder packets.
Each switch has a \emph{flow table}, that is a collection of guarded forwarding 
rules to determine the route of incoming packets. Whenever a switch receives
a packet, it checks if one of the flow table rules applies. If no rule
applies, 
the switch sends a message to the controller via a dedicated link, 
and the packet is buffered until instructions arrive. 
Depending on its policy, the controller 
instructs the switch, and possibly other switches in the network, on how to update 
their flow tables.  Such control messages between
 the controller and the switches can be processed in arbitrary
 order.

\begin{figure}[t]
\centering	
\begin{tabular}{m{.4\textwidth}m{.58\textwidth}}
\centering
\includegraphics[scale=.23]{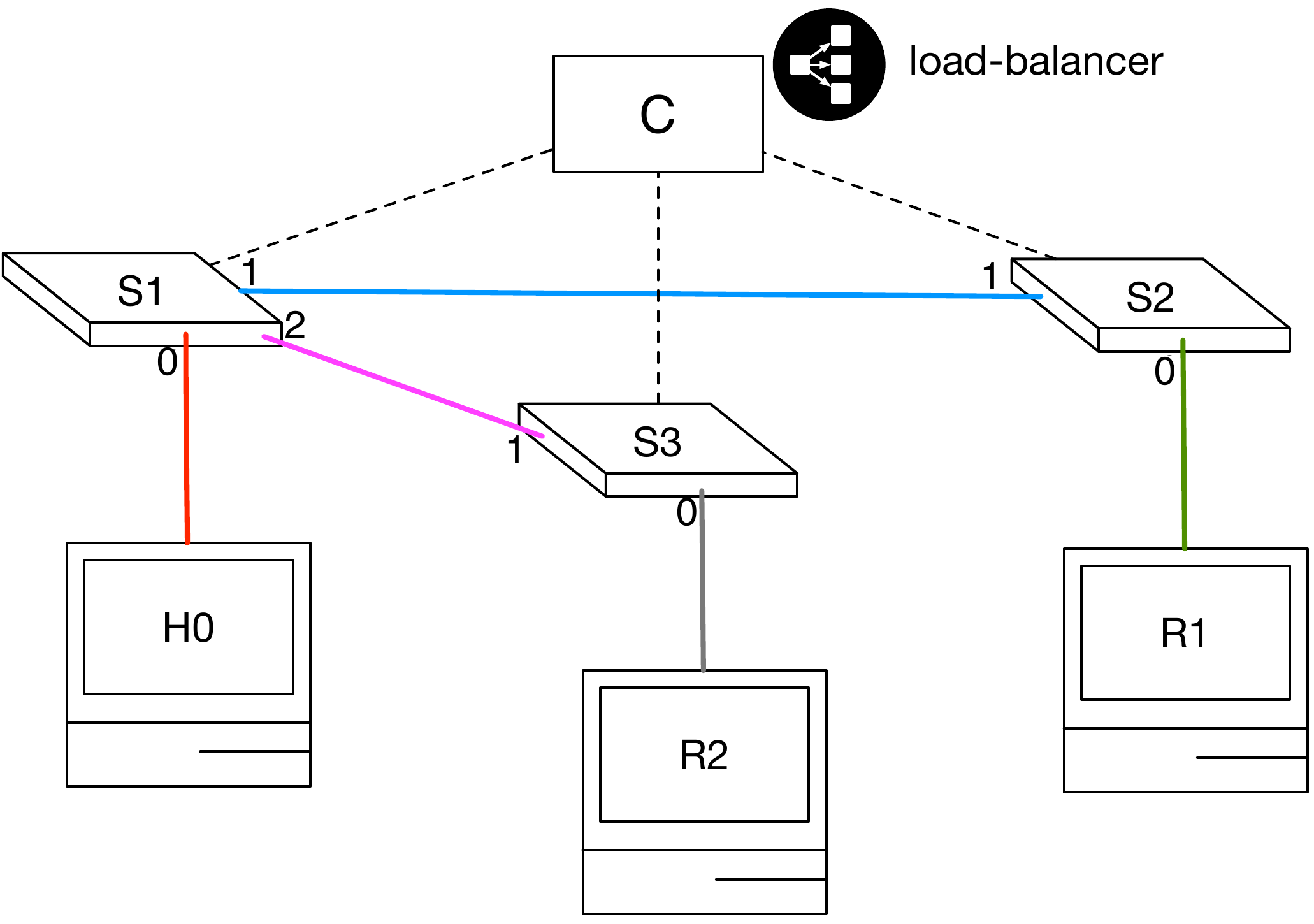}
&
\centering
\includegraphics[scale=.28]{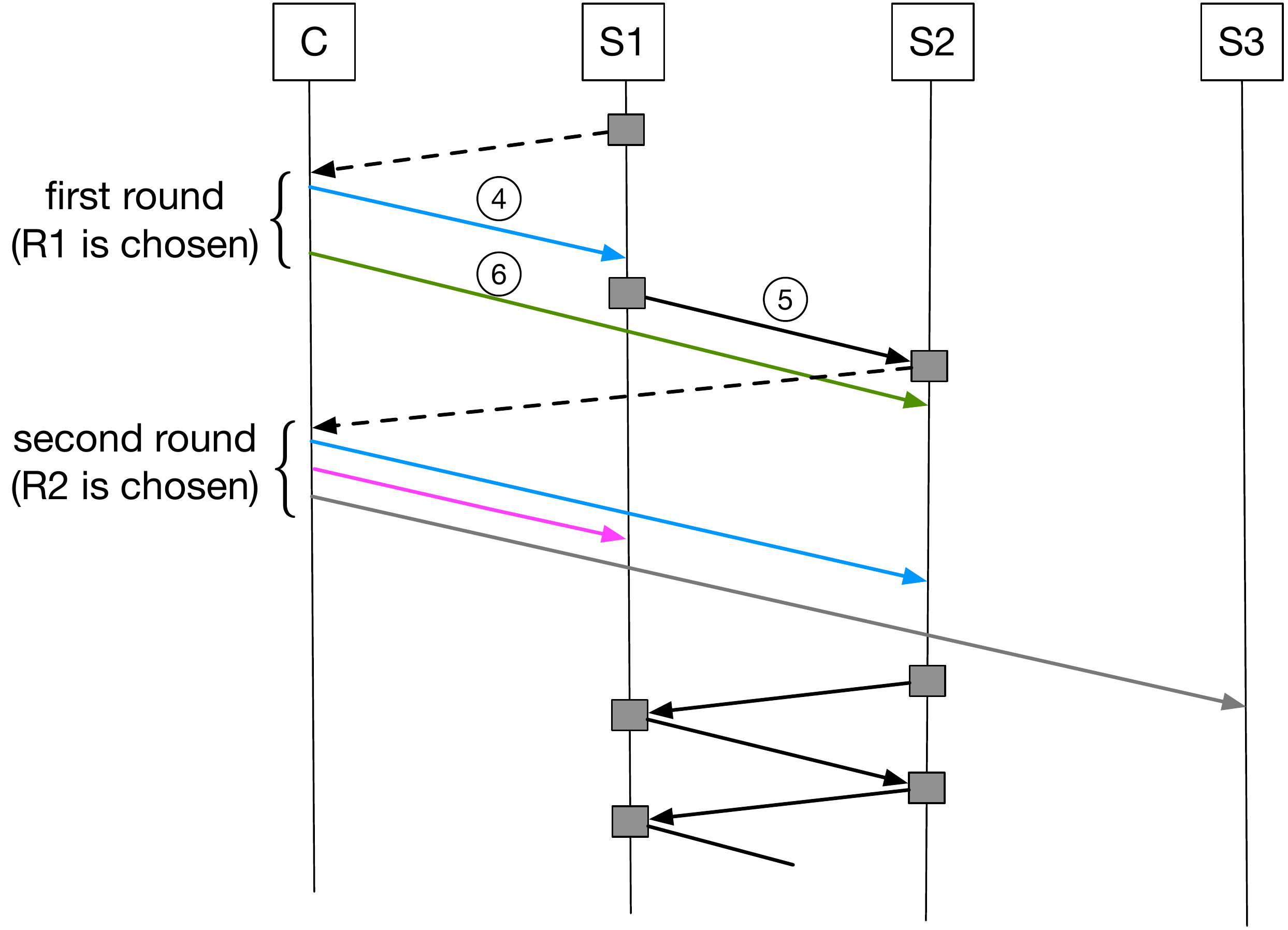}
\end{tabular}\vspace{-.3cm}\secbeg
\secbeg\secbeg
\caption{Example SDN load-balancer. On the left: structure of the
  SDN. On the right: messages exchanged in a possible execution of a
  naive controller program. Coloured arrows stand for control messages
  to switches, indicating which flow rule to install (colours specify
  the link to be used for the forwarding). Grey boxes and arrows among
  them represent packet forwardings. Dashed arrows indicate messages
  to the controller.} \vspace{-.4cm}\secbeg\secbeg\secbeg\secbeg
\label{fig:ex-sdn}
\end{figure}

We now show how a simple load-balancer can be implemented in SDN
(example taken from~\cite{DBLP:conf/pldi/El-HassanyMBVV16}) and how
potential bugs can easily arise due to the concurrent behavior and
asynchrony of message passing.  Suppose we want to balance the traffic
to a server by using two replicas \lst{R1} and \lst{R2} to which
the controller alternates the traffic in a round-robin fashion.  The
structure of the SDN is shown in \figurename~\ref{fig:ex-sdn}, on the
left: \lst{H0} is any host that wants to communicate with the
server and \lst{S1}, \lst{S2} and \lst{S3} are switches
(numbers on endpoints stand for port numbers).

Even in this simple network, an incorrect implementation of the
controller can lead to serious problems. In
Figure~\ref{fig:ex-sdn}, on the right, we show an execution of a
naive controller, which simply instructs switches to forward packets
along the shortest path to the chosen replica. This implementation
ignores the potential concurrency in actions taken by switches and
controller, leading to a forwarding loop between \lst{S1} and
\lst{S2}. In the first round, when \lst{S1} queries the controller,
\lst{R1} is chosen. The figure shows \lst{S1} forwarding the packet to
\lst{S2} before the end of the first round, i.e., before a rule is
installed on \lst{S2} (green arrow). This causes \lst{S2} to query the
controller, which triggers the second round in which the controller
chooses \lst{R2}. Thus, it sends instructions to install rules on
\lst{S2}, \lst{S1} and \lst{S3} to forward the packet to \lst{S1},
\lst{S3} and \lst{R2}, respectively. When the controller rules arrive at
\lst{S1}, it will have two contradictory instructions, telling to
forward the packet either to \lst{S2} or to
\lst{S3}. 
In the former case, the loop at the bottom of the
figure occurs. This issue can be avoided if the implementation uses
barriers---the controller will then guarantee that \lst{S2} receives
and processes control messages before taking any other action.

\secbeg
\secbeg
\secbeg

\subsection{Actor-based modelling of SDN networks}\label{sec:intro-model}

\secbeg

We now explain how we can automatically detect the above problem using
actors and model checking.
%
We use the object-oriented actor language ABS
\cite{DBLP:conf/fmco/JohnsenHSSS10,abs-tools}, where each actor type
is specified as a class, consisting of a set of fields and
methods. Actors are instances of actor classes. For instance, the
instructions:
\textsf{Controller ctrl = new Controller(); Switch s1 = new
  Switch("S1",ctrl); Host h0 = new Host("H0",s1,0);}  
%
create 3 actors: a controller \textsf{ctrl}; a switch \textsf{s1}
with name \texttt{"S1"} and a reference to \textsf{ctrl}; a host
\textsf{h0}, with name \texttt{"H0"}, connected to the switch
\textsf{s1} via the port 0.  The SDN in Figure~\ref{fig:ex-sdn} can be
modeled using one actor per component (additional data structures for
network links will be shown later).

The execution model of actors is \emph{asynchronous}. Each actor can
be thought of as a processor, with a queue of pending tasks and a
local memory. Actors are executed in parallel and, at each actor, one
task is \emph{non-deterministically} selected among all the pending
ones and executed. The syntax \textsf{Fut$<$type$>$
  f=a!m($\overline{x}$)} spawns an asynchronous task
\textsf{m($\overline{x}$)}, that is added to the queue of pending
tasks of \textsf{a}, \textsf{type} is the type of the data returned by
\textsf{m} or \textsf{Unit} if no data is returned. This task consists
in executing the method \textsf{m} of \textsf{a} with arguments
$\overline{x}$. The variable \textsf{f} is a \emph{future variable}
\cite{deboer07esop} that will allow us to check if such task has been
completed. Left-hand side of the assignment can be
omitted in case the future variable is not needed.

A partial trace of execution of our SDN actor model computed by the
model checker is (the code that the tasks below execute will be given
in Section~\ref{sec:sdn-actors:-an}):


\subsecbeg\subsecbeg   {\small \begin{align*}
	&\boxed{1\colon \textsf{h0!sendIn}} \xrightarrow{1} \boxed{2\colon\textsf{s1!switchHandlePacket}}\xrightarrow{2} \boxed{3\colon\textsf{ctrl!controlHandleMessage}} \\ \subsecbeg \subsecbeg 
	&\xrightarrow{3} \boxed{4\colon\textsf{s1!switchHandleMessage(s2)},5\colon\textsf{s1!sendOut},6\colon\textsf{s2!switchHandleMessage(r1)}} 
      \end{align*}}
\subsecbeg

Intuitively, a packet sending (\textsf{sendIn}) is executed on
\textsf{h0} (label 1), which causes the packet to be forwarded to the
switch \textsf{s1} (2), then \textsf{s1} sends a control message to
the controller (3). Finally, the controller spawns the three tasks in
the last state (parameters tell where to forward the packet).  When
executed, these tasks will produce the messages in
Figure~\ref{fig:ex-sdn} with the same numbers. Their execution
order is arbitrary: if it is the one shown in
Figure~\ref{fig:ex-sdn}, the execution trace may lead to a state
exhibiting a forwarding cycle between \textsf{s1} and \textsf{s2}. As
we will show later, this situation can be easily detected by our model
checker SYCO via an exploration of a \emph{reduced} execution tree, which
avoids equivalent executions (Section~\ref{sec:dpor-based-model}).

The ABS language provides a convenient \textsf{await} primitive that
will be used to model barriers and to rule out the behavior described
above. The instruction \textsf{await f?} synchronizes
with the termination of the task associated to the future variable
\textsf{f}, by releasing the processor (so that another task can be
scheduled) if the task is not finished. Once the awaited task is
finished, the suspended task can resume. The \textsf{await} can be
used also with boolean conditions \textsf{await b?}  to suspend the
execution of the current active task until condition \textsf{b}
holds. The formal semantics of the language can be found in
Section~\ref{semantics}.

\newcommand{\ifle}[0]{\key{if}}

\section{Semantics for SDN Networks and for Actors}\label{sec:semant-sdn-netw}

This section presents two semantics that provide the formal basis on
which we build our models: we first introduce the semantics of SDN
Networks in Section~\ref{sec:sdn-networks}, and then the semantics of
actors in Section~\ref{semantics}. The semantics of actors has been
already defined in several works (ours is a simplification of
\cite{DBLP:conf/fmco/JohnsenHSSS10}). Our formalization of the SDN
semantics is similar to that of~\cite{GuhaRF13}. We have considered a
simplification of the Openflow specification that captures the essence
of the communications of SDN networks (e.g., we have not included the
operation \emph{flood} as it behaves similarly to the considered
switch operations).

\subsection{SDN Networks}\label{sec:sdn-networks}

\begin{figure}[t]
\centering	
\includegraphics[scale=.27]{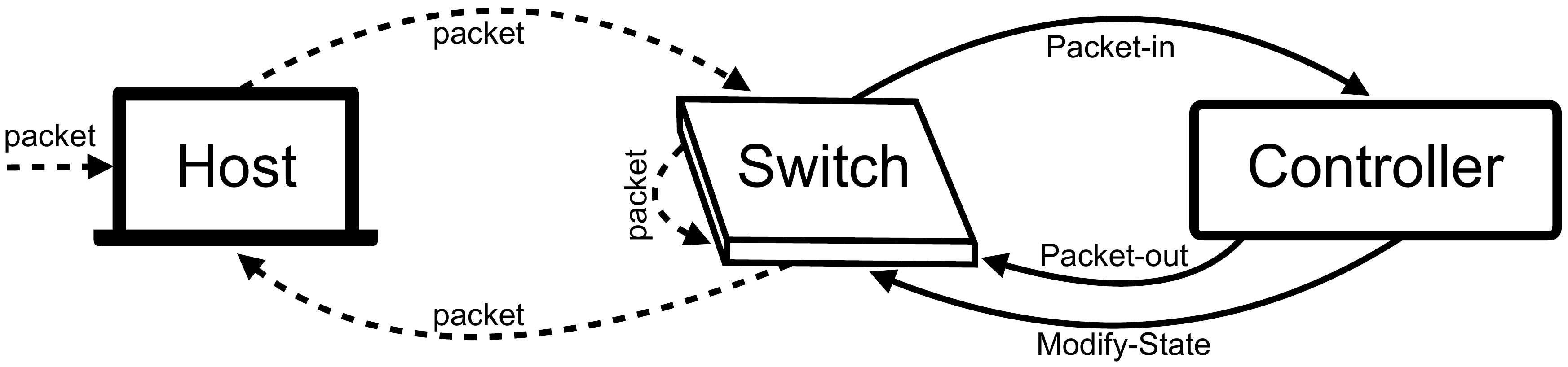}
\caption{Information flow in SDN networks} 
\label{fig:sdn-infoflow}
\end{figure}

Let us first describe the information flow of packets
and messages among the different elements in an SDN network that we
have depicted in Figure~\ref{fig:sdn-infoflow}. As in
standard networks, packets can be sent from hosts to switches and
viceversa, and also from switches to switches (see dashed arrows). The
leftmost dashed arrow represents the reception by a host of a new
packet which is fed into the network.  The specific communications of
SDN networks are performed by means of \emph{Openflow messages} (see
regular arrows), which in our simplification can be of three
types:

\begin{itemize}
\item \textsf{Packet-in}: This message is sent from a switch to the
  controller when the switch processes a packet for which it has no
  action rule to apply. The message includes the switch identifier and
  the identifier and header of the packet. The packet is buffered in
  the switch until a message of type \emph{Packet-out} is received.
\item \textsf{Modify-State}: This message is sent from the controller
  to a switch with new action rules to be inserted into the switch's
  flow-table. The message includes a flow-table entry with an action
  rule.
\item\textsf{Packet-out}: This message is sent from the controller to
  a switch to notify that it must re-try applying an action rule to a
  buffered packet. The message includes the packet header.
\end{itemize}

Figure~\ref{sdn_semantics} shows the semantics of the flow of
communications performed in our simplified SDN networks. 
%
The three types of messages below are respectively abbreviated as
\textsf{pktIn}, \textsf{modState} and \textsf{pktOut}.
\begin{itemize}
  \item A host is a
term of the form $\mathit{h(id,sid,o,in})$, where $\mathit{id}$ is the host identifier,
$\mathit{sid}$ and $o$ are, respectively, the switch identifier and port to
which the host is connected, and $in$ its input 
channel. \item A switch is of the form $\mathit{s(id,ft,b,in)}$, where $\mathit{id}$ is the
switch identifier, $\mathit{ft}$ its flow-table, $b$ its internal buffer of
packets and $\mathit{in}$ its input channel.  \item The controller is of the form
$\mathit{c(top,in)}$ where $\mathit{top}$ is the topology of the network and $in$ is its
input channel.
\item A state of the SDN network is a tuple of the form
$\langle H,S,C \rangle$ where $H = \{h ~|~ h \text{ is a host}\}$,
$S = \{s ~|~ s \text{ is a switch}\}$, that is, $H$ is a set of hosts,
$S$ is a set of switches, and, $C$ is the controller.
\end{itemize}
Letter $p$ denotes a packet. Function $header(p)$
returns its header.
Flow-tables are represented as mappings from pairs packet-header/port
to actions and are treated as a black-box through the following
functions: $\mathit{lookup(ft,\langle ph,o \rangle)}$ that returns the action
associated to the packet with header $ph$ received through port $o$ in
the flow-table $\mathit{ft}$, or $\bot$ if there is no entry for it; and,
$\mathit{put(ft,\langle ph,o \rangle,a)}$ that returns the new flow-table after
inserting in $\mathit{ft}$ the entry $\langle ph,o \rangle \mapsto a$. For
simplicity, we only consider actions of the form $send(id)$ or
$\langle send(id), o \rangle$, which indicate that the corresponding
packet should be sent, respectively, to the host $\mathit{id}$, or to the
switch with $id$ as identifier using port $o$.
%
%
Function $applyPol(top,sid,o,ph)$ represents the application of the
controller's policy using the current network topology $top$ in
result to a packet received via port $o$ with header $ph$ that the
switch with identifier $sid$ has not been able to handle. It returns a
set of pairs $\langle id,m\rangle$ where $m$ is a \textsf{modifyState}
message with an associated new flow-table entry that has to be
forwarded to the switch with identifier $id$.

A \emph{transition} or \emph{step} in the network corresponds to the
processing of a packet or message by a host, switch or the
controller. There are six (sets of) transition rules corresponding to
the different types of incoming arrows in
Figure~\ref{fig:sdn-infoflow}:

\begin{itemize}
\item \textsf{sendIn} (abbreviated as \textsc{si}): It corresponds to
  the processing by a host of a new packet which is fed into the
  network (denoted as $new(p)$), in which case the packet is forwarded
  to the switch to which the host is connected via the corresponding
  port. Note that the port is attached to the packet (denoted $o{:}p$)
  since there is only one input channel in switches.
\item \textsf{hostHandlePacket} (\textsc{hhp}): This corresponds to
  the processing by a host $h$ of a packet received from its switch,
  in which case the packet is consumed without any further action.
\item \textsf{switchHandlePacket} (\textsc{shp}): When a switch
  processes a received packet, either from a host or from another
  switch, it looks up  if there is any
  rule matching with the header of the packet and port in its flow table $\mathit{ft}$. There are three cases: (cases 1 and 2) there is a send
  action rule in the switch's flow-table, hence the packet is
  forwarded to the host (case 1) or switch (case 2) indicated in the
  action (in the latter case also the switch's port is included in the
  action); or (case 3) there is no rule for this packet, in which case
  the packet is buffered and a \textsf{Packet-in} message is sent to
  the controller.
\item \textsf{sendOut} (\textsc{so}): This corresponds to the
  processing of a \textsf{Packet-out} message by a switch. After
  looking up the header of the packet $p$ in its own flow-table $\mathit{ft}$, there are
  three cases which are analogous to those of
  \textsf{switchHandlePacket} except that the packet is in the
  switch's buffer (instead of in its input channel), and that if no
  action rule is found in the switch's flow-table the packet is
  dropped.
\item \textsf{switchHandleMessage} (\textsc{shm}): It corresponds to
  the processing of a \textsf{Modify-State} message by a switch, in
  which case the received action rule is inserted into the switch's
  flow-table.
\item \textsf{controlHandleMessage} (\textsc{chm}): The controller
  receives a \textsf{Packet-in} message from a switch $s$ in result to
  a packet that the switch $s$ has not been able to handle. As a
  result, the controller sends a set $\mathit{ms}$ of \textsf{Modify-State} messages
  with new action rules to a selected set of switches (as specified by
  the controller's policy with the current network topology), and a
  \textsf{Packet-out} message to switch $s$.
\end{itemize}

A derivation $\ex \equiv
\state_0 \rightarrow \cdots \rightarrow \state_n$ is {\em complete}
if $\state_0$ is the initial state and $\state_n = \langle H,S,C
\rangle$ is the final state such that every
message and packet in their channels has been processed (their input
channels are empty).
 We use $exec(S)$ to denote the set of
all possible executions starting at state $S$.

\newcommand{\rrrule}[2]{
\!
\begin{array}{c}
#1 \\
\hline\\[-0.25cm] 
#2 
\end{array}
\vspace*{0.0cm}
}

\begin{figure}[bhp]

\resizebox{.85\textwidth}{!}
{
\begin{mathpar}

\begin{aligned}

&\inferrule*[leftskip=4em,left=(si)]
 {h = h(id,sid,o,in \cup \{new(p)\}) \quad s = \mathit{s(sid,ft,b,in')}}
 {\langle \{h\} \cup H, \{s\} \cup S, C\rangle \rightarrow
   \langle \{h(id,sid,o,in)\} \cup H, \{\mathit{s(sid,ft,b,in' \cup \{o{:}p\})}\} \cup S, C\rangle}

\\[2ex]

&\inferrule*[leftskip=4em,left=(hhp)]
 {h = h(id,sid,o,in \cup \{p\})}
 {\langle \{h\} \cup H, S, C\rangle \rightarrow
   \langle \{h(id,sid,o,in)\} \cup H, S, C\rangle}

\\[2ex]

&\inferrule*[leftskip=4em,left=(shp$_1$)]
 {s = \mathit{s(sid,ft,b,in \cup \{o{:}p\})} \quad h = h(id,sid,o',in') \\ send(id) = \mathit{lookup(ft,\langle header(p){,}o\rangle)}}
 {\langle \{h\} \cup H, \{s\} \cup S, C\rangle \rightarrow
   \langle \{h(id,sid,o',in' \cup \{p\})\} \cup H, \{\mathit{s(sid,ft,b,in)}\} \cup S, C\rangle}

\\[2ex]

&\inferrule*[leftskip=4em,left=(shp$_2$)]
 {s = \mathit{s(sid,ft,b,in \cup \{o{:}p\})} \quad s' = \mathit{s(sid',ft',b',in')}\\
   send(sid'{,}o') = \mathit{lookup(ft,\langle header(p){,}o\rangle)}}
 {\langle H, \{s,s'\} \cup S, C\rangle \rightarrow
   \langle H, \{\mathit{s(sid,ft,b,in)}, \mathit{s(sid',ft',b',in' \cup \{o'{:}p\})}\} \cup S, C\rangle}

\\[2ex]

&\inferrule*[leftskip=4em,left=(shp$_3$)]
 {s{=}\mathit{s(sid,ft,b,in \cup \{o{:}p\})} \quad s' = \mathit{s(sid,ft,b \cup \{o{:}p\},in)} \quad
 \bot = \mathit{lookup(ft,\langle header(p){,}o\rangle)}}
 {\langle H, \{s\} \cup S, c(top,in')\rangle \rightarrow
   \langle H, \{s'\} \cup S, c(top,in' \cup \{\mathsf{pktIn(sid,o,id(p),header(p))}\})\rangle}

\\[2ex]

&\inferrule*[leftskip=4em,left=(so$_1$)]
 {s = \mathit{s(sid,ft,b \cup \{o{:}p\},in \cup \{\mathsf{pktOut(ph)}\})}\\ ph=header(p) \quad h = h(id,sid,o',in') \quad
  send(id) = \mathit{lookup(ft,\langle header(p){,}o\rangle)}}
 {\langle \{h\} \cup H, \{s\} \cup S, C\rangle \rightarrow
   \langle \{h(id,sid,o,in' \cup \{p\})\} \cup H, \{\mathit{s(sid,ft,b,in)}\} \cup S, C\rangle}

\\[2ex]

&\inferrule*[leftskip=4em,left=(so$_2$)]
 {s = \mathit{s(sid,ft,b \cup \{o{:}p\},in \cup \{\mathsf{pktOut(ph)}\})}\\ ph=header(p) \quad  s' = \mathit{s(sid',ft',b',in')} \quad 
  send(sid',o') = \mathit{lookup(ft,\langle header(p){,}o\rangle)}}
 {\langle H, \{s,s'\} \cup S, C\rangle \rightarrow
   \langle H, \{\mathit{s(sid,ft,b,in)}, \mathit{s(sid',ft',b',in' \cup \{o'{:}p\})}\} \cup S, C\rangle}

\\[2ex]

&\inferrule*[leftskip=4em,left=(so$_3$)]
 {s = \mathit{s(sid,ft,b \cup \{o{:}p\}, in \cup \{\mathsf{pktOut(ph)}\})} \\ \mathit{ph=header(p)} \quad
  \bot{=}\mathit{lookup(ft,\langle header(p){,}o\rangle)}}
 {\langle H, \{s\} \cup S, C\rangle \rightarrow
   \langle H, \{\mathit{s(sid,ft,b,in)}\} \cup S, C\rangle}

\\[4ex]

&\inferrule*[leftskip=4em,left=(shm)]
 {s = \mathit{s(sid,ft,b,in \cup \{\mathsf{modState(\langle ph,o\rangle \mapsto a)}\})}}
 {\langle H, \{s\} \cup S, C\rangle \rightarrow
   \langle H, \{\mathit{s(sid,put(ft,\langle ph,o\rangle,a),b,in)}\} \cup S, C\rangle}

\\[4ex]

&\inferrule*[leftskip=4em,left=(chm)]
 {c = c(top,cin \cup \{\mathsf{pktIn(sid,o,pid,ph)}\}) \quad
  s = \mathit{s(sid,ft,b,sin)} \\ ms = \mathit{applyPol(top,sid,o,ph)} \quad
  ms_{id} = \{m ~|~ \langle id,m\rangle \in ms \} \\ 
  S' = \{\mathit{s(sid',ft',b',in')} \mid \mathit{s(sid',ft',b',in)} \in S,
  in' = in \cup ms_{sid'}\} 
  }
 {\langle H, S \cup \{s\}, c\rangle \rightarrow
   \langle H, S' \cup \mathit{s(sid,ft,b,sin \cup ms_{sid} \cup \{\mathsf{pktOut(ph)}\})}, c(top,cin)\rangle}

\end{aligned}

\end{mathpar}
}
\caption{Semantics of SDN networks}\label{sdn_semantics}
\end{figure}

\subsection{Syntax and Semantics for Actor Programs}\label{semantics}

The grammar below describes the syntax of the language ABS in which
SDN models will be defined:

\medskip

\(
\begin{array}{lll}
  
  P & {:}{:}{=} & M~\bar{C}\\
  C & {:}{:}{=} & \textbf{class}~c(\bar{T}~\bar{x})\{\bar{M}\}\\
 M & {:}{:}{=} & T~m(\bar{T}~\bar{x})\{s;\}\\
 s & {:}{:}{=}  & s~;~s \mid x=e \mid 
 \ifle~\mathit{b}~\iftethen~s~\ifteelse~s  \mid
                  \while~\mathit{b}~\whilebody~\mathit{s} \mid  m({\bar{z}}) \\
 &&  \mid  x=\newb~C(\bar{y})   \mid f=x!m({\bar{z}})
    \mid \await{f}~ \mid \await{b}\\
\end{array} 
\)

\medskip

\noindent
Here, $x$, $y$, $z$ denote variables names, $f$ a future variable
name, and $s$ a sequence of instructions.  For any entity $A$, the
notation $\bar{A}$ is used as a shorthand for $A_1,...,A_n$.  We use
the special identifier \texttt{this} to denote the current
actor.  For generality, the syntax of expressions $e$,
Boolean conditions $b$ and types $T$ is left unspecified. As in the
object-oriented paradigm, a class denotes a type of actors including
their behavior, and it is defined as a set of fields and
methods. Lastly, $m({\bar{z}})$ denotes standard (synchronous)
method calls, which are only allowed on the actor itself, whereas
``$!$'' is used for asynchronous method calls (see
Section~\ref{sec:intro-model}).

\begin{figure}[t]

\resizebox{.85\textwidth}{!}{
\begin{mathpar}  
\begin{alignedat}{3}
&\inferrule*[leftskip=3em,left=(mstep)]{
\buffer(\objid,\bot,\heap,\queue)=\selectObject(\state) \\
  \task(\tkid,m,\tv,s) = \selectTask(\buffer(\objid,\bot,\heap,\queue)) \quad s  \not = \epsilon \quad   \state  \microstepstar{\objid\cdot\tkid} \state'   
 }
 {\state \macrostep{} \state'} 

\\[1ex]

&\inferrule*[leftskip=3em,left=(asy)]
{\taskId=\task(\tkid,m,l, \futvar{x}=y~!~m_1(\many{z});s) \quad
  \objid_1=\tv(y)\quad
  \tkid_1=\text{fresh}() \quad
  l_1{=}{\it newlocals}(\bar{z},m_1,l)}
 {\buffer(\objid,\tkid, \heap,\queue \cup \{\taskId\})\sepo
 \buffer(\objid_1,\tkid',\heap', \queue')\rrderiv\\ 
\buffer(\objid,\tkid, \heap,  \queue {\cup} \{\task(\tkid,m,l[\futvar{x}{\mapsto}\tkid_1],s)\})\sepo \buffer(\objid_1,\tkid',\heap',  \queue' {\cup} \{\task(\tkid_1,m_1,\tv_1,\body{m_1})\})}

\\[1ex]

&\inferrule*[leftskip=3em,left=(syn)]
{\taskId=\task(\tkid,m,l,m_1(\many{z});s) \quad l_1{=}{\it newlocals}(\bar{z},m_1,l)}
 { \textsc{(syn)}~\buffer(\objid,\tkid \quad
   \heap,\queue \cup \{\taskId\})\sepo \rrderiv
\buffer(\objid,\tkid,\heap,  \queue {\cup} \{\task(\tkid,m,\tv_1,\body{m_1};s)\})}

\\[1ex]

&\inferrule*[leftskip=3em,left=(new)]
{\taskId=\task(\tkid,m,l,{x=\key{new}~D(\bar{y});s}) \quad \objid_1{=}\text{fresh}()\\
h'=\newheap(D) \quad
l'=l[x\rightarrow  \objid_1]\quad
\key{class}~D(\bar{f})\{\ldots\}}
 {\buffer(\objid,\tkid,h, \queue  \cup  \{\taskId \} )
\rrderiv 
{\buffer(\objid,\tkid,h, \queue  \cup \{\task(\tkid,m,l',s)\}) \sepo
  \buffer(\objid_1,\bot,h'[\bar{f} \mapsto \tv(\bar{y})],\emptyset)}}

\\[1ex]

&\inferrule*[leftskip=3em,left=(await)$_1$]
{\taskId = \task(\tkid,m,\tv,\key{await}~\futvar{x};s) \quad   
     \tv(\futvar{x})=\tkid_1 \quad
     \task(\tkid_1,m_1,\tv_1, \epsilon) \in \state}
  {\buffer(\objid,\tkid, \heap, \queue \cup \{\taskId\}) 
  \rrderiv \buffer(\objid,\tkid, \heap,\queue \cup \{\task(\tkid,m,\tv,s)\})} 

\\[1ex]

&\inferrule*[leftskip=3em,left=(await)$_2$]
{\taskId=\task(\tkid,m,\tv,\key{await}~\futvar{x};s) \quad
    \tv(\futvar{x})=\tkid_1 \quad 
    \task(\tkid_1,m_1,\tv_1,\epsilon) \not \in \state}
  {\buffer(\objid,\tkid, \heap, \queue \cup \{\taskId\}) 
  \rrderiv  \buffer(\objid,\bot,\heap,  \queue \cup \{\task(\tkid,m,\tv,\key{await}~\futvar{x};s)\}) } 
 
\\[1ex]
  
&\inferrule*[leftskip=3em,left=(return)]
{\taskId =\task(\tkid,m,\tv,\epsilon)}
 {\buffer(\objid,\tkid,\heap,\queue \cup \{\taskId\}) 
 \rrderiv  
 \buffer(\objid,\bot,\heap, \queue\cup \{\taskId\})}

\end{alignedat}
\end{mathpar}
}
 \caption{Semantics of concurrent primitives of actor programs}\label{Summarized_semantics_await}
\end{figure}

Figure~\ref{Summarized_semantics_await} presents the semantics of the
actor model. An \emph{actor} is a term of the form
$\buffer(\objid,\tkid,\heap,\queue)$, where $\objid$ is the actor
identifier, $\tkid$ is the identifier of the \emph{active task} that
holds the actor's lock or $\bot$ if the actor's lock is free, $\heap$
is its local heap and $\queue$ is the queue of tasks in the actor. A
{\em heap} $\heap$ is a mapping
$\heap :\Fields(C) \mapsto \mathbb{V}$, 
where 
$\mathbb{V}$
stands for the set of references and values.
A \emph{task}
$\taskId$ is a term $\task(\tkid,m,l,s)$ where $\tkid$ is a unique
task identifier, $m$ is the method name executing in the task, $\tv$
is a mapping from local variables to $\mathbb{V}$, and $s$ is the
sequence of instructions to be executed. Finally, a
  global state $S$ is a set of actors.
As actors do not share their states, the semantics can be
presented as a macro-step semantics \cite{DBLP:conf/fase/SenA06}
(defined by means of the transition ``$\macrostep{}$'') in which the
evaluation of all statements of a task takes place serially (without
interleaving with any other task) until it gets to a \emph{release
  point}, i.e., a point in which the actor's processor becomes idle
due to the return or an await instruction. In this case, rule {\sc (mstep)} is
applied to select an available task from an actor, namely relation
$\selectObject(\state)$ is applied to select \emph{non-deterministically} an actor
$\buffer(\objid,\bot,h,\queue)$ in the state with a non-empty queue $\queue$, and,
$\selectTask(\buffer(\objid,\bot,h,\queue))$ to select \emph{non-deterministically} a task of
$\queue$. 
Micro-step transitions are written $\rrderiv$ and define
evaluations in task \tkid by actor \objid within a given macro-step.
As before, the sequential instructions are standard and thus omitted.
In {\sc (new)}, an active task $\taskId$ in actor
$\objid$ creates a new actor of class
$D$ with a fresh identifier
$\objid_1=\text{fresh}()$, which is introduced to the state with a
free lock. Here
$h'=\newheap(D)$ stands for a default initialization on the fields of
class $D$. Rule {\sc (syn)} simply replaces in task
$\tkid$ the statement with the method call to
$m_1$ by its body. Rule {\sc (asy)} spawns a new task (the initial
state is created by $\it
newlocals$) with a fresh task identifier
$\tkid_1$ which is stored in the future variable
$\futvar{x}$.  We assume $\objid \neq \objid_1$, but the case $\objid
= \objid_1$ is analogous, the new task
$\tkid_1$ is simply added to the queue $\queue'$ of actor
$\objid_1$.  In rule {\sc
  (await)$_1$}, the future variable
$\futvar{x}$ we are awaiting for points to a finished task and thus
the \key{await} can be completed. The finished task identified with
$\tkid_1$ is looked up in all actors in the current state (written as
$\task(\tkid_1,m_1,\tv_1, \epsilon) \in
\state$).  Otherwise, {\sc
  (await)$_2$} yields the lock so that any other task of the same
actor can take it. The behaviour of {\sc await} on Boolean conditions
is analogous.  When rule {\sc (return)} is executed, the task is
\emph{finished}, but it remains in the queue so that rules {\sc
  (await)$_1$} and {\sc (await)$_2$} can be applied.
A derivation $\ex \equiv
\state_0 \macrostep{} \cdots \macrostep{} \state_n$ is {\em complete}
if $\state_0$ is the initial state and all actors in $\state_n$ are of
the form $\buffer(\objid,\bot,\heap,\queue)$, where for all $\taskId
\in \queue$ it holds that $\taskId \equiv
\task(\tkid,m,\tv,\epsilon)$. We use $exec(S)$ to denote the set of
all possible executions starting at state $S$.


\secbeg
\secbeg
\secbeg
\section{SDN-Actors: an actor based encoding of SDN programs 
}\label{sec:sdn-actors:-an}
\secbeg
\secbeg

We present the concept of \emph{SDN-Actor} in 3 steps:
Section~\ref{sec:network-topology} describes the creation and
initialization of the actors according to the
topology. Section~\ref{sec:switches} provides the encoding of the
operations and communication for \lst{Switch} and \lst{Host} actors.
Section~\ref{sec:controller} proposes the encoding of the
controller. Altogether, our encoding provides an actor-based semantics
foundation of SDN networks that follow the OpenFlow specification
\cite{openflow} captured by the semantics in
Section~\ref{sec:sdn-networks}.





\secbeg\secbeg\secbeg\vspace{-.2cm}

\subsection{Network topology}\label{sec:network-topology}
\noindent
The topology can be given as a relation with two types of links:
\begin{enumerate}
\item \emph{SHlink(s,h,o)}:  switch $s$ is connected to host
  $h$ through the port $o$
\item \emph{SSlink(s$_1$,i$_1$,s$_2$,i$_2$)}: switch $s_1$ is connected
  via port $i_1$ to port $i_2$ ofs $s_2$ 
\end{enumerate}\secbeg
from which we automatically generate the initial
configuration as follows.
\begin{definition}[initial configuration]\label{def:conf}
  Let $S$ and $H$ be, respectively, the set of different switch and
  host identifiers available in the \emph{link} relations that define
  the network topology. The initial configuration (method \lst{init\_conf}) is defined as:
\begin{itemize}
\item We create a controller actor \emph{\lst{Controller ctrl=new
      Controller()}}
\item For each $\emph{\lst{sid}}{\in}S$, we create an actor  \emph{\lst{Switch s=new Switch(sid,ctrl)}}
\item For each $\emph{\lst{hid}}{\in}H$, we create an actor \emph{\lst{Host
      h=new Host(hid,s,o)}} where \emph{\lst{s}} is the reference to
  the switch actor, \emph{\lst{o}} the port identifier, that \emph{\lst{hid}}
 is
  connected to.
\item The data structures \emph{\lst{srefs}} and \emph{\lst{hrefs}}
  store, respectively, the relations between identifier in the topology and
  reference in the program, for all switches in $S$ and hosts in $H$.
  \item The data structure  \emph{\lst{ntw}} contains the link relations
    in the network topology.
  \item The synchronous call
    \emph{\lst{ctrl.addConfig(srefs,hrefs,ntw)}} initializes in the
    controller the topology relations and the references to switches
    and hosts such that the controller can send control messages to
    redirect the traffic to the involved links.
\end{itemize}
\secbeg
\end{definition}
\secbeg
\secbeg
\secbeg
\subsecbeg
\begin{example}\label{ex:network-topology-1}
By applying Definition~\ref{def:conf} to
the topology in Figure~\ref{fig:ex-sdn}, given as the relation:
{\footnotesize $SHlink(S1,H0,0)$, $SHlink(S2,R1,0)$, $SHlink(S3,R2,0)$},
{\footnotesize $SSlink(S1,1,S2,1)$}, and {\footnotesize $SSlink(S1,2,S3,1)$}, we obtain the following
initial configuration which constitutes the \lst{init\_conf} method from
which the execution starts:
%
%
\secbeg\subsecbeg
\begin{lstlisting}[name=basicexs,framextopmargin=11pt]
init_conf() { Controller ctrl = new Controller(); Switch s1 = new Switch("S1",ctrl);
         Switch s2 = new Switch("S2",ctrl); Switch s3 = new Switch("S3",ctrl);
         Host h0 = new Host("H0",s1,0); Host r1 = new Host("R1",s2,0); 
         Host r2 = new Host("R2",s3,0); 
         $Map$<SwitchId,Switch> srefs = {"S1":s1, "S2":s2, "S3":s3};
         $Map$<HostId,Host> hrefs = {"H0":h0, "R1":r1, "R2":r2};
         $List$<Link> ntw = [SHLink("S1","H0",0), SSLink("S1",1,"S2",1),..];
         $\label{conf:last}$ctrl.addConfig(srefs,hrefs,ntw);  } 
\end{lstlisting}
\secbeg\subsecbeg The data structures \lst{srefs} and \lst{hrefs} are
implemented using maps, and the network \lst{ntw} as a heterogeneous list.  The
use of data structures is nevertheless orthogonal to the
encoding as actors.
We just assume standard functions to create, initialize, access them 
(like getters, put, take, lookup, etc.) that will appear in italics in
the code.
\end{example}

 
 

\subsection{The switch and host classes}\label{sec:switches}
\begin{figure}[p]
\begin{center}
\begin{tabular}{|l|}
\hline
\begin{lstlisting}[name=basicexs]
type SwitchId=...  type HostId=... type PortId=...  type PacketId=...
type PacketH=...   type Packet=... type Action=...  type Link=...
type MatchF=(PacketH,PortId);
\end{lstlisting} \\\hline

\begin{lstlisting}[name=basicexs]
class Host(HostId hid, Switch s, PortId o) {
 $\label{sendin}$Unit sendIn(Packet p){
    s!switchHandlePacket(p,o);
 }
 $\label{hhp}$Unit hostHandlePacket(Packet p){
    $/*output~packet*/$
 }
}
class Switch(SwitchId sid, Controller ctrl) {
  $\label{ft}$ $Map$<MatchF,Action> flowT={};
  $\label{buf}$ $Map$<PacketId,(Packet,PortId)> buffer={};
 Unit switchHandlePacket(Packet p, PortId o){
   $\label{shp1}$ Action l=$lookup$(flowT,($getHeader$(p),o));
   $\label{shp2}$ if ($isSwitch$(l))
        $getSwitch$(l)!switchHandlePacket(p,$getPort$(l));
   $\label{shp3}$ else if ($isHost$(l))
        $getHost$(l)!hostHandlePacket(p);
        else {
        $\label{shp4}$buffer=$put$(buffer,$getId$(p),(p,o));
        $\label{shp5}$ctrl!controlHandleMessage(sid,o,$getId$(p),$getHeader$(p));
    }
 }
 Unit sendOut(PacketId pi){
   $\label{send1}$ Packet p; PortId o;
   (p,o)=$take$(buffer,pi);
   $\label{send2}$ Action l=$lookup$(flowT,($getHeader$(p),o));
   $\label{send3}$ if ($isSwitch$(l))
        $getSwitch(l)$!switchHandlePacket(p,$getPort$(l));
   $\label{send4}$ else if ($isHost$(l))
        $getHost(l)$!hostHandlePacket(p);
   $\label{send5}$ $/*else~packet~is~dropped*/$
 }
 Unit switchHandleMessage(MatchF m, Action a){
   flowT=$put$(flowT,m,a);
 }
}
\end{lstlisting} \\\hline
\end{tabular}
\end{center} \vspace{-.4cm}\secbeg \secbeg \secbeg\subsecbeg
\caption{Type declarations (top) and actor-based host and switch
  classes (bottom)}
\label{class-switch} \secbeg \secbeg\subsecbeg\subsecbeg\subsecbeg \vspace{-.4cm}
\end{figure}

Figure~\ref{class-switch} presents the actor-based \lst{Switch} and \lst{Host}
classes. 
We include at the top some \lst{type} declarations that are assumed and must be implemented (such as
identifiers, packets and their headers, etc.). There are two main data
structures implemented in more detail to make explicit the information
they contain:
\secbeg
\begin{itemize}
\item the \lst{buffer} at Line~\ref{buf} (L\ref{buf} for short) is a \emph{map} that must contain pairs of
  packet and input port indexed by their \lst{PacketId}.
\item the flow table \lst{flowT} (L\ref{ft}) is implemented as a \emph{map} indexed by the
  so-called \emph{match field} \cite{openflow} represented by type
  \lst{MatchF} in Figure~\ref{class-switch}. The match field is composed
  by information stored in the header of a \lst{Packet} (retrieved by
  function $\mathit getHeader$) and the input port. For a given
  matching, the flow table contains the \lst{Action} the switch has to
  perform upon the reception of the \lst{Packet}. An action \lst{l}
  can be of three types: i) send the packet to a host \lst{h}, ii)
  send the packet to the port \lst{o} of a switch \lst{s}, iii) drop
  the packet.  Given an action \lst{l}, function $\mathit isSwitch$
  respectively $\mathit isHost$ succeeds if the action is of type ii)
  respectively i), and functions $\mathit getSwitch$, $\mathit getHost$ and
  $\mathit getPort$ return the \lst{s}, \lst{h} and \lst{o}
  respectively. The full implementation must allow duplicate entries
  (non-deterministically selected), and the use of
  wildcards in the match fields, but these aspects are unrelated to the
  encoding of SDN actors, and  skipped for simplicity.
\end{itemize}\subsecbeg\subsecbeg
Upon creation, hosts receive their identifier and a reference to the switch 
and the port identifier they are connected to (defined as class parameters that are initialized 
at the actor creation). Their method \lst{sendIn} is used to send a
packet to the switch, and method \lst{hostHandlePacket} to receive a
packet from the switch.
Switches receive upon creation their identifier and a reference to the
controller. They have as additional fields: (a) the flow table \lst{flowT} (as
described above) in which they store the actions to take upon
receiving each kind of package, and (b) a buffer in which they
store packets that are waiting for a response from the controller.
Switches can perform three operations: (1) \lst{switchHandlePacket}
receives a packet, looks up in the flow table the action to be made
L\ref{shp1}, and, if there is an entry for the packet in the table, it
asynchronously makes the corresponding action (either send it to a
host L\ref{shp3} or to a switch L\ref{shp2}).  Otherwise, it sends a
\lst{controlHandleMessage} request and puts the packet and input port
in the buffer (L\ref{shp4} and L\ref{shp5}) until it can be handled
later upon receipt of a \lst{sendOut}; (2) \lst{sendOut} receives a
packet identifier that corresponds to a waiting packet, retrieves it
from the buffer (L\ref{send1}), looks up the action \lst{l} to be
performed in the flow table, and makes the corresponding asynchronous
call (as in \lst{switchHandlePacket}); (3) \lst{switchHandleMessage}
corresponds to a message received from the controller with an
instruction to update the flow table. Other switch operations like
\emph{forward packet}, that is similar to \lst{sendOut} but directly tells
the switch the action to be performed, or \emph{flood}, that sends a
packet through all ports except the input port, can be encoded
similarly and are used in the 
experiments in Section~\ref{sec:verif-sdn-prop}.

\begin{example}\label{ex:sendin}
  In \lst{init\_conf}, after L\ref{conf:last}, we add \lst{h0\!sendIn(p)},
  where \lst{p} is a packet to be sent to the IP address of the
  replica servers (the information on the destination is part of the
  packet header). This is the only asynchronous task that \lst{init\_conf}
  spawns. Its execution in turn spawns a new task
  \lst{s1\!switchHandlePacket(p,0)} at L\ref{sendin}, that does not
  find an entry in \lst{flowT} at L\ref{shp1} and spawns a
  \lst{controlHandleMessage} task on the controller at L\ref{shp5},
  whose code is presented in the next section.
\end{example}

\secbeg
\secbeg
\secbeg
\secbeg
\subsection{The controller}\label{sec:controller}
\secbeg

\begin{figure}[t]
\begin{center}
\begin{tabular}{|l|}
\hline
\begin{lstlisting}[name=basicexs]
class Controller() {
  $Map$<SwitchId,Switch> srefs={};
  $Map$<HostId,Host> href={};
  $List$<Link> ntw=[];
  Unit addConfig($Map$<SwitchId,Switch> sr, $Map$<HostId,Host> hr, $List$<Link> n){
       $\mathit{/*references~to~switches~and~hosts~and~network~topology~initialized*/}$
  }
  Unit controlHandleMessage(SwitchId sid, PortId o, PacketId p, PacketH h){
      $\label{actions}$$List$<(SwitchId,MatchF,Action)> l=$applyPolicy$(sid,o,h);
     $\label{while1}$ while (not($isEmpty$(l))) {
         $\label{shm1}$ SwitchId s1; Action a1; MatchF m1;
         (s1,m1,a1)=$head$(l);
         $\label{shmb1}$ $lookup$(srefs,s1)!switchHandleMessage(m1,a1);
         $\label{next}$ l=$tail$(l);
      }
      $\label{sob}$ $lookup$(srefs,sid)!sendOut(p);
   }
 }
\end{lstlisting} \\\hline
\end{tabular}
\end{center}\vspace{-.4cm}\secbeg\secbeg
\secbeg
\secbeg
\caption{Controller class (without  barriers)}\secbeg\secbeg
\secbeg
\label{class-controller}
\end{figure}

After creating the controller actor, the method \lst{addConfig} is
invoked synchronously to initialize the references to switches and
hosts and set up the initial network topology (see L\ref{conf:last}).
A simple controller is presented in Figure~\ref{class-controller}.
When a switch asynchronously invokes \lst{controlHandleMessage}, the
controller applies the current policy---function
$applyPolicy$ must be implemented for each different type of
controller. The implementation of the policy typically requires the
definition of new data structures in the controller to store
additional information (see Section~\ref{sec:verif-sdn-prop}).
When applying the policy for a given \lst{SwitchId}, \lst{PortId} and
\lst{PacketH}, we obtain a list of switch identifiers and
corresponding actions to be applied to them (as a data-structure of
type
\lst{$List$<(SwitchId,MatchF,Action)>}). The while loop at
L\ref{while1} in \lst{controlHandleMessage} asynchronously invokes
\lst{switchHandleMessage} at L\ref{shmb1} on each of the switches in
the list, and passes as parameter the corresponding action to be
applied for the given match entry. Finally, it notifies at L\ref{sob}
the switch from which the packet came that this can be
sent out.  More sophisticated controllers that build upon this
encoding are described in Section~\ref{sec:verif-sdn-prop}.

\begin{example}\label{ex:applypolicy}
  In the example, \lst{applyPolicy} corresponds to the load-balancer
  described in Section~\ref{sec:overview}, which directs external
  requests to a chosen replica in a round-robin fashion. For the call
  $applyPolicy$\lst{(s1,0,h)}, it chooses \lst{r1} and thus, it
  returns in L\ref{actions} two actions: \lst{(s1$\rightarrow$s2)},
  \lst{(s2$\rightarrow$r1)}, i.e., one action to install in \lst{s1}
  the rule to send the packet to \lst{s2}, and the second to install
  in \lst{s2} the rule to send it to \lst{r1}. For simplicity, we
  assume that the \lst{Action} just contains the location to which the
  packet has to be sent (without including the port). The while loop
  thus spawns two asynchronous calls,
  \lst{s1\!switchHandleMessage(m1,s2)} and
  \lst{s2\!switchHandleMessage(m1,r1)}. Besides, it sends a
  \lst{s1\!sendOut(p)} in L\ref{sob}. Several problems may arise in
  this implementation. One problem, as explained in
  Section~\ref{sec:overview}, is that the packet is sent from \lst{s1}
  to \lst{s2} before the control message is processed by
  \lst{s2}. Then, \lst{s2} gets the packet and it does not find any
  matching rule, thus it sends a \lst{controlHandleMessage} to the
  controller. Applying the above policy, the controller chooses now as
  replica \lst{r2} and returns the actions: \lst{(s2$\rightarrow$s1)},
  \lst{(s1$\rightarrow$s3)}, \lst{(s3$\rightarrow$r2)}, i.e., the
  packet should be sent to \lst{r2} by first sending from \lst{s2} to
  \lst{s1} (first action), and so on.  This might create the
  circularity depicted in Figure~\ref{fig:ex-sdn}.

\end{example}

\subsection{Soundness of the Encoding}
An execution in the
network is characterized by the messages in the queues of the
switches, hosts, and controller and the state of their data
structures.
First of all, let us define the equivalence between an input channel with its buffer $(in,b)$ and a queue of pending tasks with its
buffer $(\queue,\text{buffer})$. Let us notice here that even though
we have used different notation for $b$ and $\text{buffer}$, we use $b =
\text{buffer}$ to denote the equality of information in both structures, that
is, they have exactly the same packets and with the same ports.

\begin{definition}
An input channel $in$ with a buffer of pending packets $\text{b}$ and a queue of pending tasks
$\queue$ with a buffer of pending packets $\text{buffer}$ are equivalent, written $(in,\text{b}) \equiv (\queue,\text{buffer})$ if and only if:
\begin{enumerate}
\item $in = \emptyset = \queue$ and $b = \text{buffer}$ or
\item otherwise on the following holds:
\begin{description}
\item[pktOut:] 
$in = \{\textsf{pktOut(ph)}\} \cup in', ~\exists \task, p$ such that $
\queue = \{\task(\_,\textsf{sendOut},l,\_)\} \cup \queue', \text{b}
= \text{b'} \cup \{o{:}p\}$, $\text{buffer}
= \text{buffer}' \cup \{(p,o)\}$, $getId(p) = l[\textsf{pi}]$ and 
$getHeader(p) = \textsf{ph}$, and $(in',b') \equiv
(\queue',\text{buffer'})$.
\item[modState:] $in = \{\textsf{modState}(\langle \textsf{ph,o}\rangle \mapsto \textsf{a})\} \cup in', \exists\task$ such
that $\small \queue = \queue' \cup \\\{\task(\_,\textsf{switchHandleMessage},l,\_)\}$,
$(\langle\textsf{ph,o}\rangle \mapsto a) = (l[\textsf{m}] \mapsto l[\textsf{a}])$ and $(in',b) \equiv (\queue',\text{buffer})$,
\item[pktIn:] $in = \{\textsf{pktIn(sid,o,pid,ph)}\} \cup in',
~ \exists \task$ such that $\queue
= \queue' \cup \\\{\task(\_,\textsf{controlHandleMessage},l,\_)\},$ $~sid = l[\textsf{sid}],~ pid
= l[\textsf{p}],~ph = l[\textsf{h}],~o = l[\textsf{o}]$ and $(in',b) \equiv (\queue',\text{buffer})$. 
\item[packet:] $in =  \{o{:}p\} \cup in', ~\exists \task$
such that $\queue = \{\task(\_,\textsf{switchHandlePacket},l,\_)\} \cup \queue'$, 
$~o = l[\textsf{o}], ~ p = l[\textsf{p}]$ and $(in',b) \equiv (\queue',\text{buffer})$.
\item[packet-out:] $in =  \{p\} \cup in', ~\exists \task$
such that $\queue
= \{\task(\_,\textsf{hostHandlePacket},l,\_)\} \cup \queue',~p = l[\textsf{p}]$, and 
$(in',b) \equiv (\queue',\text{buffer})$.
\item[packet-in:] $in =  \{new(pkt)\} \cup in', ~\exists \task$
such that $\queue = \{\task(\_,\textsf{sendIn},l,\_)\} \cup \queue',~ pkt = l[\textsf{p}]$, and $(in',b) \equiv (\queue',\text{buffer})$.
\end{description}
\end{enumerate}
\end{definition}
Now, we can define the equivalence between an SDN state and an
SDN-Actor state.
\begin{definition}[equivalence]
An SDN state $S = \langle H,Sw,C \rangle$ and an SDN-actor state $S_a$
are equivalent, written $S \equiv S_a$, if and only
if:
\begin{description}
\item[Host:] $\forall h(id,sid,o,in) \in H, \exists! a(\_,\_,h,\queue) \in S_a$
such that $(in,\emptyset) \equiv (\queue,\emptyset)$, \\ $id = h[hid],
sid = h[s],$ and $ o = h[o]$.
\item[Switch:] $\forall s(id,ft,b,in) \in Sw, \exists! a(\_,\_,h,\queue) \in S_a$
such that $(in,b) \equiv (\queue,h[\text{buffer}])$, $id = h[sid],$
and $ft =  h[flowT]$.
\item[Controller:] $C = c(top,cin)$ and $\exists !a(id,\_,h,\queue) \in S_a$
such that \\ $(cin,\emptyset) \equiv (\queue,\emptyset)$, $related(top,\{h[\text{srefs}],h[\text{href}],h[ntw]\})$, \\ and $\forall a(\_,\_,h',\_) \in
S_a$, $id= h'[ctrl]$.
\end{description}
\end{definition}
Let us notice here that we use
$\mathit{related(top,\{h[srefs],h[href],h[ntw]\})}$ to clarify that information
about the topology is coherent in both the controller and the
controller actor.


The following
theorem ensures the soundness of our modelling. Essentially we
guarantee that, for a given SDN network that follows the OpenFlow
specification, any execution in the network has an equivalent
execution in the SDN-Actor model. The proof can be found in the
appendix. We denote as $S_a^{ini}$ the SDN-Actor state defined in
Definition \ref{def:conf}, i.e., after executing method
\lst{init\_conf()} and all asynchronous calls
  to method \lst{sendIn} containing the packets to be delivered. Furthermore,
$S^{ini} \equiv S_a^{ini}$.

\begin{theorem}\label{th:modelling}
Let $S^{ini}$ and $S_a^{ini}$ be an SDN state and an SDN-Actor state,
respectively. 
\begin{enumerate}
\item For every execution $S^{ini} \rightarrow S^1 \rightarrow... \rightarrow
S^n \in exec(S^{ini}), \exists
S^{ini}_a \macrostep{} ... \macrostep{} S_a^n \in
exec(S_{a}^{ini})$ such that $S^n \equiv S_a^n$.
\item For every execution $S_a^{ini} \macrostep{} S_a^1 \macrostep{}... \macrostep{}
S_a^n \in exec(S_a^{ini}), \exists
S^{ini} \rightarrow ... \rightarrow S^n \in
exec(S^{ini})$ such that $S^n \equiv S_a^n$.
\end{enumerate}
\end{theorem}

\section{Implementing barriers using conditional synchronization}\label{sec:barriers}

\begin{figure}[p]
\begin{center}
\begin{tabular}{|l|}
\hline
\begin{lstlisting}[name=basicexs]
class Controller() {
  $Map$<SwitchId,Switch> srefs={};
  $Map$<HostId,Host> href={};
  $List$<Link> ntw=[];
  $\label{b1}$@\textcolor{blue}{$Map$$<$SwitchId,List$<$Fut$<$Unit$>>$ barrierMap=\{\};}@
  $\label{b2}$@\textcolor{blue}{$Set$$<$SwitchId$>$ barrierOn = $\emptyset$;}@
  Unit addConfig($Map$<SwitchId,Switch> sr, $Map$<HostId,Host> hr, $List$<Link> n){
       $\mathit{/*references~to~switches~and~hosts~and~network~topology~initialized*/}$
  }
  Unit controlHandleMessage(SwitchId sid, PortId o, PacketId p, PacketH h){
      $\label{actionsb}$$List$<(SwitchId,MatchF,Action)> l=$applyPolicy$(sid,o,h);
      $\label{iniswitchlist}$@\textcolor{blue}{$List$$<$SwitchId$>$ ls = [];}@
     $\label{while1b}$ while (not($isEmpty$(l))) {
         $\label{shm1b}$ SwitchId s1; Action a1; MatchF m1;
          (s1,m1,a1)=$head$(l);
         $\label{barrieroffcall1}$ @\textcolor{blue}{barrierWait(s1);}@
         $\label{shmb1}$ @\textcolor{blue}{Fut$<$Unit$>$
  f=}@$lookup$(srefs,s1)!switchHandleMessage(m1,a1); $\label{barriermap}$
          @\textcolor{blue}{barrierMap=$putAdd$(barrierMap,s1,f);}\label{barrierput}@
          @\textcolor{blue}{ls = $add$(ls,s1);}@
         $\label{nextb}$ l=$tail$(l);
      }
     $\label{whilex}$ @\textcolor{blue}{\textbf{while}
  (not($isEmpty$(ls))) \{}@
          $\label{barrierofRequest}$@\textcolor{blue}{barrierWait($head$(ls));}@
          $\label{barrierofRequest2}$@\textcolor{blue}{barrierRequest($head$(ls));}@
          $\label{AfterBarrierRequest}$@\textcolor{blue}{ls=$tail$(ls);}@
      @\textcolor{blue}{\}}@
     $\label{barrierofcall2}$ @\textcolor{blue}{barrierWait(sid);}@
     $\label{sobb}$ @\textcolor{blue}{Fut$<$Unit$>$
  f=}@$lookup$(srefs,sid)!sendOut(p);
     $\label{barrierput2}$ @\textcolor{blue}{barrierMap=$putAdd$(barrierMap,sid,f);}@
  }
  Unit barrierWait (SwitchId sid){
       $\label{barrieroff1}$ await not($contains$(barrierOn,sid))?;
  }
  Unit barrierRequest (SwitchId sid){
     $\label{barrierfalse}$ barrierOn=$add$(barrierOn,sid);
     $\label{futsid}$ List<Fut<Unit>> futSid=$take$(barrierMap,sid);
     $\label{while}$ while (not($isEmpty$(futSid)) {
           $\label{futsid2}$ Fut<Unit> fi=$head$(futSid);
           $\label{await2}$  await fi?;
           $\label{futsidnext}$ futSid=$tail$(futSid);
     }
    $\label{barriertrue}$ barrierOn=$delete$(barrierOn,sid);
  }
}
\end{lstlisting} \\\hline
\end{tabular}
\end{center}\vspace{-.4cm}\secbeg\secbeg
\secbeg
\secbeg
\caption{Extension of Controller class with barriers}\secbeg\secbeg
\secbeg
\label{class-controller-barriers}
\end{figure}

Barriers \cite{openflow} have been designed to force a switch to
handle previous control messages, and thus avoid problems such as the
one described above.

\begin{definition}[OF barrier]\label{def:barriers-1}
  Following OpenFlow \cite{openflow}, upon receipt of a \emph{barrier
    message}, the switch must finish processing all
  previously-received controller messages, before executing any
  messages received after the \emph{barrier message}.
\end{definition}




Figure~\ref{class-controller-barriers} shows our modelling that
intuitively consists in the controller not sending further messages to
any switch on which a barrier has been activated, until this switch
acknowledges that all previous control messages have been already
processed.  The main points in the implementation are:
\begin{enumerate}
\item The controller creates a future variable at L\ref{shmb1} for
  every asynchronous task that it posts on all switches.
\item it keeps in \lst{barrierMap} the list of future variables (not
  yet acknowledged) for each of the switches (\lst{putAdd} in
  L\ref{barriermap} adds the future variable to the list indexed by
  \lst{s1} in the map).
\item The controller keeps in \lst{barrierOn} the set of switches with
  an active barrier.
\item A barrier on a switch consists in the controller awaiting on the
  list of future variables that the switch needs to acknowledge to
  ensure that its control messages have already been processed (method
  \lst{barrierRequest}).
\item All control messages must be now preceded by a call to
  \lst{barrierWait} that checks if the corresponding switch has an
  active barrier, L\ref{barrieroff1}. This is because while suspended
  in a barrier, the controller can start to process another
  \lst{controlHandleMessage} unrelated to the previous one, but which
  affects (some of) the same switches for which a barrier was set. So,
  we cannot send messages to them until their barriers are set to
  off. Similarly, the call to \lst{barrierRequest} must also be
  preceded by a call to \lst{barrierWait} since
  \lst{barrierRequest} is indeed modelling the send to the switch of a
  control message (the \emph{barrier message}).
\end{enumerate}

Note that this is not a restriction on the type of controllers we
model, but rather an effective way to encode barriers using actors and
conditional synchronization (by means of the \lst{await} instructions)
that ensures the behaviour of OpenFlow barriers.

%

The next theorem states that our implementation of barriers via
methods \lst{barrierRequest} and \lst{barrierWait} provide a sound
encoding of the OF \emph{barrier} messages in
Definition~\ref{def:barriers-1}.

\begin{theorem}[soundness of barriers] 
  Given any state $S$ in any execution of the SDN-Actor model right
  before executing L\ref{barrierofRequest2} with switch $sid$ as
  parameter (i.e., the state before activating a barrier over $sid$),
  and the state $S'$ right before executing L\ref{AfterBarrierRequest}
  (i.e., the state after receiving the acknowledgement of the
  barrier), the following holds:

  \begin{itemize}
  \item All \lst{switchHandleMessage} and \lst{sendOut} tasks in the
    queue of switch $sid$ in state $S$ have been completely executed
    in state $S'$.
  \item No \lst{switchHandleMessage} nor \lst{sendOut} task have been
    spawned over switch $sid$ in any middle state between $S$ and
    $S'$.
  \item No other \lst{barrierRequest} call for switch $sid$ is
    performed between $S$ and $S'$.
  \end{itemize}
\end{theorem}

\begin{proof}
  Let us firstly define an invariant which holds for every possible
  state $S''$ of any execution of the SDN-Actor model:
$$\forall a(sid,\_,\_,\queue) \in
S'' \mbox{~and~} \forall\task(tk,m,\_,\_) \in \queue,~ m \in \{\mbox{\lst{switchHandleMessage}},\mbox{\lst{sendOut}}\}$$
$$ \exists!
a(cid,\_,h,\_) \in S''\mbox{~such that~} \mathit{tk} \in
h[\mbox{\lst{barrierMap}}][\mbox{\lst{sid}}]$$
The invariant states that every spawned \lst{switchHandleMessage} or
\lst{sendOut} task $tk$ on a switch $sid$ is recorded by means of a
future variable in the list associated to $sid$ in the
\lst{barrierMap} field of the controller (i.e.
$\mathit{tk} \in h[\mbox{\lst{barrierMap}}][\mbox{\lst{sid}}]$). Note the abuse of
notation $\in$ to check existence of an element in a $List$
data-structure, and $[]$ to access the value of a key in a $Map$
data-structure. It can be seen that after making any asynchronous call
to method \lst{switchHandleMessage} (L\ref{barriermap}) or
\lst{sendOut} (L\ref{sobb}), the corresponding future variable is
always recorded in the \lst{barrierMap} field (L\ref{barrierput} and
L\ref{barrierput2}).

Now, given the controller of the state $S$,
$a(cid,\_,h_c,\queue_c) \in S$, for every task
$\task(tid,\mbox{\lst{controlHandleMessage}},l_c,\mbox{\lst{barrierRequest(l)}}{;s})
\in \queue_c$, we have a derivation
$S = S_0 \microstepstar{cid.tid} S_1 \macrostep{} ... \macrostep{} S_n
\macrostep{cid.tid} S_{n+1} = S'$ such that $S_1$ is the global state
after executing the micro-step transitions of such task until it stops
at L\ref{await2}, and $S_{n+1}$ is the first state where
$h_c[\mbox{\lst{barrierOn}}]$ does not contain the switch
$sw = l[\mbox{\lst{sid}}]$. Let us notice that if such stop is not
performed, then every task in $sw$ has already finished and
$\mbox{\lst{barrierRequest}}$ is performed in a single macro-step
($S_0 = S_{n}$).
Then, we know that
$ \forall i \in \{0,...,n+1\}, \exists a(sw,\_,\_,\queue_i) \in S_i$
with $\queue_i = \mathit{SHP_i \cup SO_i \cup SHM_i}$ such that:

\begin{itemize}
\item $\mathit{SHP}_i$ contains all \lst{switchHandlePacket} tasks,
\item $\mathit{SO}_i$ contains all \lst{sendOut} tasks, and,
\item $\mathit{SHM}_i$ contains all \lst{switchHandleMessage} tasks.
\end{itemize}

By the definitions of the states $S_1$ and $S_{n+1}$, we know that
$\forall i \in \{1,...,n\}$, $sw \in h_c[\mbox{\lst{barrierOn}}]$.
Hence, $\forall i \in \{1,...,n\}$, the condition of the \lst{await}
instruction at L\ref{barrieroff1} does not hold, thus, the task is
suspended in state $S_i$, and, consequently, no
\lst{switchHandleMessage} nor \lst{sendOut} task can be spawned in any
state $S_i$. Therefore,
$\forall i \in \{1,...,n\}, \forall j \in \{i,...,n\}, \mathit{SHM}_j$
(resp. $\mathit{SO}_j$) never contains more tasks than
$\mathit{SHM}_i$ (resp. $\mathit{SO}_i$).
Similarly, no other call to $\mbox{\lst{barrierRequest}}$ can be
performed due to the call to \lst{barrierWait} in
L\ref{barrierofRequest}, which implies that there cannot be two active
barriers over the same switch.

Finally, since $sw$ no longer belongs to \lst{barrierOn} in $S_{n+1}$,
we know that
$\forall \mathit{fut} \in
h_c[\mbox{\lst{barrierMap}}][\mbox{\lst{sid}}]$, the task
$l_c[\mathit{fut}]$ has finished, since for each variable
$\mathit{fut}$, the \lst{await} statement in L\ref{await2} has
succeeded. Moreover, using the invariant, we know that all the tasks
in $\mathit{SHM_n}$ and $\mathit{SO}_n$ have their corresponding
future variable in
$\mathit{h_c[\mbox{\lst{barrierMap}}][\mbox{\lst{sid}}]}$, and
therefore all of them have finished.

\end{proof}

\secbeg
\secbeg

\secbeg
\secbeg
\section{DPOR-based model checking of SDN-Actors}\label{sec:dpor-based-model}
\secbeg

Model checking tools deal with a combinatorial blow-up of the state
space (a.k.a. the state space explosion problem) that must be faced to solve
real-world problems. This problem
is exacerbated in the context of SDN programs, because of the concurrent and distributed nature of
networks: all network components (switches, hosts, controllers) are
distributed nodes that run in parallel and whose concurrent tasks can
interact. As we have seen, a controller message sent from a switch can
change the state of another switch, and affect the route of an
incoming packet. Thus, a model checker needs to explore all possible
reorderings of \emph{dependent} tasks (i.e., those whose execution
might interfere with each other) leading to a huge number of possible
executions even for networks with a low number of nodes and packets.
Additionally, the state space is unbounded because hosts may
generate unboundedly many packets that could be simultaneously
traversing the network. 

There are two \emph{incomplete} approaches to handle unbounded inputs:
one is to impose a bound $k$ on the number of packets of each type (as
e.g. in\cite{DBLP:conf/nsdi/CaniniVPKR12}) and the other one is to use
abstraction (as e.g. in \cite{DBLP:conf/fmcad/MajumdarTW14}). In the
former, the search space is exhausted for the considered input, but
there could be bugs that only show up when more packets are
considered.  In the latter, abstraction requires to lose information
and bugs may only show up when the omitted information is
considered. 
Therefore, the sources of incompleteness are different, and the approaches can
complement each other.  Our tool SYCO uses the
former, e.g., in Example~\ref{ex:sendin} we have considered one packet
(limit $k=1$).  The rest of the section presents the key features of
our approach
assuming such a $k$  bound.

\secbeg
\secbeg
\secbeg

\subsection{DPOR-based model checking in actors}\label{sec:cont-sens-dpor}
\secbeg

DPOR \cite{DBLP:conf/popl/FlanaganG05} is able to dynamically identify
and avoid the exploration of redundant executions and prune the
search space exponentially. It is based on the idea of initially
exploring an arbitrary interleaving of the various concurrent tasks,
and \emph{dynamically} tracking dependent interactions between them to
identify backtracking points where alternative paths in the state
space need to be explored. Two tasks are \emph{independent} when
changing their order of execution will not affect their combined
effect.  
When DPOR is applied to actor systems, there are inherent
reductions 
\cite{DBLP:conf/forte/TasharofiKLLMA12-short} because: (i) we can
atomically execute each task (without re-orderings) until a return or
an \lst{await} instruction are found, as concurrency is
non-preemptive and the active task cannot be interrupted. This avoids
having to consider the reorderings at the level of instructions (as
one must do in thread-based concurrency), and allows us to work at the
level of tasks.
(ii) Also,
two tasks can have a dependency only if they belong to
the same actor. This is because only the actor itself can modify its
private memory.  \newcommand*\circled[1]{\tikz[baseline=(char.base)]{
    \node[shape=circle,draw,inner sep=1](char) {#1};}} \secbeg
\begin{example}
  Figure~\ref{ex:tree} shows the search tree computed by DPOR for our
  SDN-Actor program without barriers. 
  \begin{figure}[t]\secbeg
\secbeg
\secbeg
\secbeg
\secbeg
\secbeg
\begin{center}
\fbox{
\includegraphics[width=0.9\textwidth]{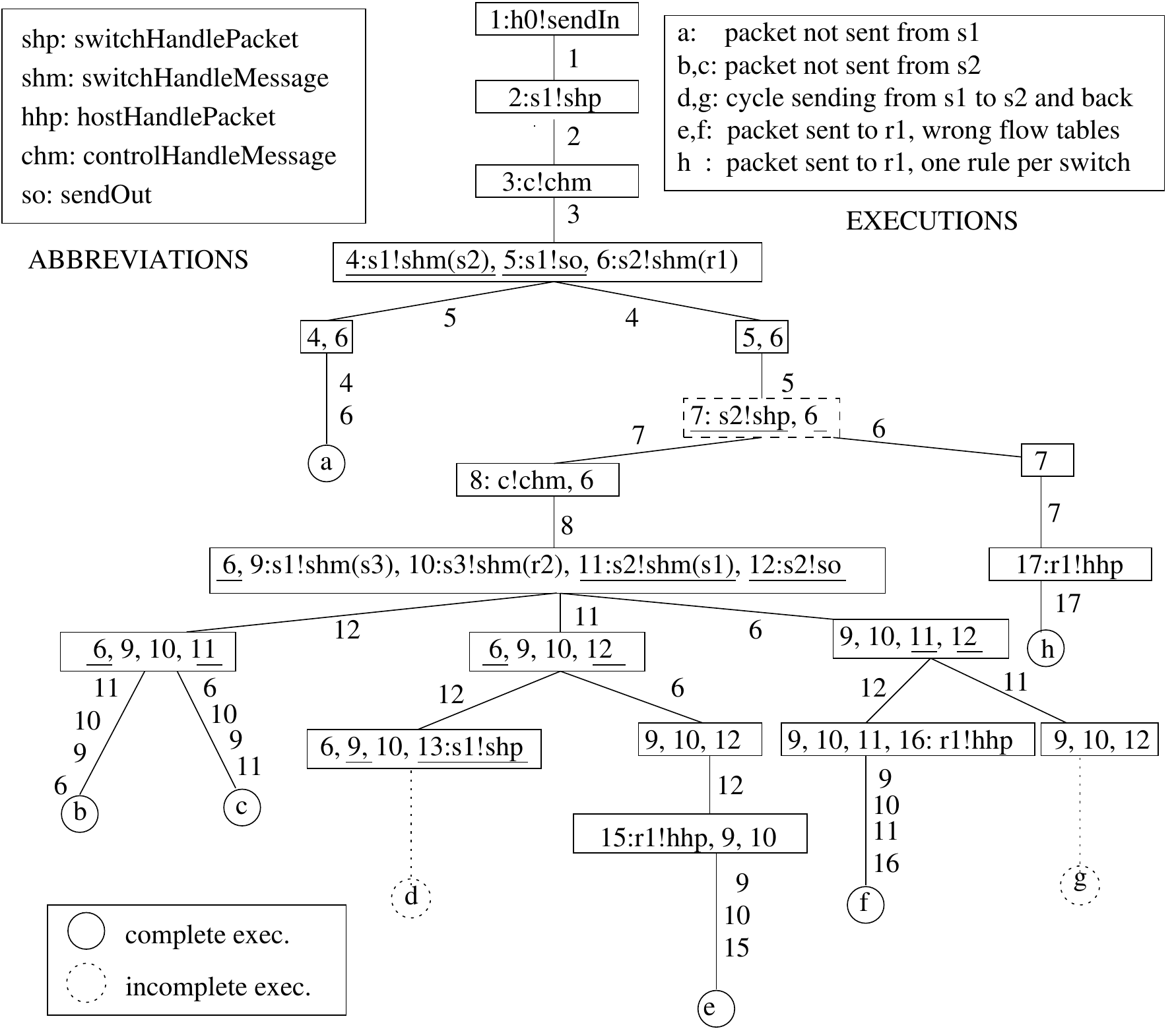}
}
\secbeg
\secbeg
\caption{Search tree for running example w/o barriers (rightmost
  branch w/ barriers)} \label{ex:tree}
\end{center} \vspace{-.5cm}
\secbeg
\secbeg
\secbeg
\secbeg
\secbeg
\secbeg
\end{figure}
  It has no redundancy, i.e., each
  execution corresponds to a different behavior on the packet arrival
  and/or the actions installed in the flow tables (see top right
  descriptions). At each node (i.e., state), we show the available
  tasks. A task is given an identifier the first time it appears, and
  afterwards only its identifier is shown. Method names are
  abbreviated as shown in the top left, and parameters are
  omitted except in tasks executing \lst{switchHandleMessage}, for
  which we only include the switch identifier that is part of the
  \lst{Action} to be installed. For instance,
  \textsf{4:s1!shm(s2)} is a task with identifier \lst{4}, that
  will execute method \lst{switchHandleMessage} on \lst{s1} and will
  add to its flow table the information that the packet must be sent
  to \lst{s2}. Labels on the edges show the task(s) that have been
  executed. At each state, we underline the tasks which have an
  interacting dependency. The execution starts by executing the
  \lst{init\_conf} method in Example~\ref{ex:network-topology-1} with the
  instruction \lst{sendIn} added in Example~\ref{ex:sendin} which appears
  in the root. The next two steps have one task available, but in the
  fourth state we have tasks \lst{4} and \lst{5}, belonging to the
  same actor, whose reordering needs to be considered (leading to
  branching), while \lst{6} is independent of them.  Out of the 8
  branches of the tree, only the rightmost execution \circled{h}
  corresponds to the correct behavior in which the packet is actually
  sent to \lst{r1} and the actions are installed in the flow tables in
  the expected order. In execution \circled{a} the packet does not
  arrive at the destination because the \lst{sendOut} is executed
  before the action has been installed. Executions \circled{d} and
  \circled{g} correspond to the cycle described in
  Section~\ref{sec:overview}, each of them with different installations
  of actions.
\end{example}
\subsecbeg Importantly, we do not need specific optimizations to use
the DPOR algorithm in \cite{DBLP:conf/cav/AlbertABGS17} to model check
SDN-Actors. The use of \lst{await} (is already covered by DPOR and)
does not require any change either and, as
expected, 
the search tree for the implementation with barriers only contains
branch \circled{h}. The difference arises from task \lst{3} in the
tree: in the presence of barriers, this leads to a state in which we
have asynchronous calls \lst{4} and \lst{6} and task \lst{3}
suspended at the \lst{await} in L\ref{await2} (awaiting for the
termination of \lst{4} and then of \lst{6}). Therefore, the
dependent tasks \lst{4} and \lst{5} will not coexist because \lst{5}
is not spawned until \lst{4} and \lst{6} terminate.


\secbeg
\secbeg
\subsection{Entry-level and context-sensitive independence}\label{sec:entry-level-context}
\secbeg

When two tasks that belong to the same actor are found, in the context of DPOR techniques,
independence is commonly over-approximated by requiring that actor fields accessed by
one task are not modified by the other.
%
%
In our model, all tasks posted on a given switch access its flow table,
namely \lst{sendOut} and \lst{switchHandlePacket} read it and
\lst{switchHandleMessage} writes it. Thus, in principle, any task
executing \lst{switchHandleMessage} is considered dependent on the
other two. This explains the tasks underlinings in the figure and the
branching in the tree. 
When there are multiple packets traversing the network usually
different packets access distinct entries in the
flow table. This results in the inaccurate detection of many
dependencies hence producing redundant executions.  Using Constrained
DPOR~\cite{DBLP:conf/cav/AlbertGIR18}, we alleviate this state
space explosion:
\begin{enumerate} \secbeg
\item \emph{Entry-level independence}. 
  We adopt a finer-grained notion of \emph{entry-level independence}
 for which an access to entry $i$ is independent from an access to
 $j$ if $i \neq j$.  
 This aspect is not visible when considering a single
 packet as in the example, as all accesses to the flow table refer to
 the same entry. However, by simply adding another packet to the
 erroneous program, the state explosion is huge and the system times
 out if entry-level independence is not implemented, while it computes
 92 executions (exploring 761 states) with entry-level independence.

\item \emph{Context-sensitiveness}. Even when two tasks $t$ and $p$
  access the same entry, Constrained DPOR
  \cite{DBLP:conf/cav/AlbertGIR18} introduces some further checks that
  avoid redundant explorations. If the state before executing both
  tasks satisfies a certain \emph{independence annotation}, then the
  executions of $p$ and $q$ are guaranteed to commute. Hence, one of
  the derivations can be pruned and further exploration from it is
  avoided.  For instance, executing two consecutive
  \lst{switchHandleMessage} on the same entry might lead to the same
  state if the flow table contains duplicate entries, as our
  implementation allows. An example of independence annotation for
  these two tasks is the check of duplicate entries in the state.

\end{enumerate}
Although entry-level independence in theory could be proved
automatically by using SMT solvers
(see~\cite{DBLP:conf/cav/AlbertGIR18}), this is not yet possible in
our system, and we have declared annotations which are valid for any
SDN model. Let us explain the most representative annotations for method
\lst{switchHandleMessage(m,a)}: 
\begin{enumerate} \item \lst{indep(switchHandlePacket(pi,pk),\!matchHead\&Port(getHeader(pk),pi,m{)}{)}}
  denotes that tasks executing \lst{switchHandleMessage(m,a)} are
  independent of those executing \lst{switchHandlePacket(pi,pk)} if
  the matched field of the message does not match the header and the
  input port of the packet (the condition is checked by the auxiliary
  function \lst{matchHead\&Port}).

\item
\lst{indep(switchHandleMessage(m2,a2),indepSwitchMsgeMsge(m,a,m2,a2))}\ \ 
denotes that tasks executing \lst{switchHandleMessage(m,a)} are
independent of those executing \lst{switchHandleMessage(m2,a2)} if
the matched fields \lst{m} and \lst{m2} are independent (they do not match with
the same entries in the flow table), but actions \lst{a} and
\lst{a2} are equals (the condition is checked by the auxiliary function
\lst{indepSwitchMsgeMsge}).
\end{enumerate}

\secbeg
\secbeg
\secbeg
\secbeg \vspace{-.2cm}
\subsection{Comparison of DPOR reductions with related work}\label{sec:dpor-reductions-sdn}
\secbeg

Other model checkers for SDN programs have used DPOR-based algorithms
before
\cite{DBLP:conf/nsdi/CaniniVPKR12,DBLP:conf/fmcad/MajumdarTW14}. According
to the experiments in the NICE tool, DPOR only achieves a 20\%
reduction of the search space because even the finest granularity does
not distinguish independent flows. The reason for this modest
reduction might be that it does not take advantage of the inherent
independence of the code executed by the distributed elements of the
network (switches, host, clients), nor to the fact that barriers allow
removing dependencies, as our actor-based SDN model does.
In Kuai \cite{DBLP:conf/fmcad/MajumdarTW14}, a number of optimizations
are defined to take advantage of these aspects. Such optimizations
must be (1) identified and formalized in the semantics, (2) proven
correct and, (3) implemented in the model checker. Instead, due to our
formalization using actors, the optimizations are already implicit in
the model and handled by the model checker without requiring any
extension. 
Another main difference with Kuai is that they make two important
simplifications to the kind of SDNs they can handle: (i) they assume a
simplified model of switches in which a switch gets suspended (i.e.,
does not process further packets nor controller messages) while
awaiting a controller request. The error showed in
Example~\ref{fig:ex-sdn} would thus not be captured. We do not make any
simplification and thus
a switch can start to process a new packet while awaiting the
controller and can also receive other controller actions (triggered by
other switches). (ii) It works on a class of SDNs in which the size of
the controller queue is one. Therefore, it will not capture potential
errors that arise due to the reordering of messages by the controller.
In contrast, our model checker works on the general model of SDN
networks.

\secbeg \secbeg \vspace{-.2cm}
\section{Implementation and experimental evaluation}
\label{sec:verif-sdn-prop}
\secbeg

This section describes how to use our model checking tool and its
visualization capabilities in Section~\ref{sec:model-checking-tool},
and then the experimental evaluation carried out on a series of
standard SDN benchmarks in Section~\ref{sec:check-sdn-prop}.

\subsection{The model checking tool and its visualization capabilities}\label{sec:model-checking-tool}
 
We have built an extension for property model checking on top of the
SYCO tool \cite{DBLP:conf/cc/AlbertGI16}. It can be used through an
online web interface available at:
\small\url{http://costa.fdi.ucm.es/syco} \normalsize by selecting
the POR algorithm \texttt{CDPOR} and disabling the automatic
generation of independence constraints. All benchmarks we are
describing in this section can be found in the folder
\textsf{JSS19}.  
In order to run the model checker, the user first
opens one of these benchmarks and clicks over the button
\textsf{Apply}. By default SYCO makes a full exploration of the
execution. However, by using the \textsf{Settings}, it is possible to
change the default options. In particular, by  selecting
\textsf{Property checking}, the exploration finishes after finding an
execution trace that violates the property being checked.  In order to
define the property $P$ under test, we add to the controller a new
method called \texttt{error\_message} and encode $P$ as a Boolean
function $F_p$ using the programming language itself. Then, in all
places where the property has to hold, we add an \texttt{if} statement
checking the negation of $F_p$ and if it holds we call asynchronously
to \texttt{error\_message} on the
controller. 
Then property holds for the given input if and
only if there is no trace in the execution tree including a call to
\texttt{error\_message}.

%
The result of executing the model checker is shown in the console at
the bottom, where \syco first prints the number of executions explored
and the output state for each explored execution.
The output state contains the actors modelling the controller, the
switches and the hosts created during the execution. Each actor is
represented as a term with three arguments: the actor identifier, the
actor type or class, and the final values of their fields. 

\subsection{Checking SDN properties in case studies}\label{sec:check-sdn-prop}

 To evaluate our approach, we have implemented a
series of standard SDN benchmarks used in previous work
\cite{DBLP:conf/fmcad/MajumdarTW14,DBLP:conf/pldi/BallBGIKSSV14,DBLP:conf/pldi/El-HassanyMBVV16}.
Our goal is on the one hand to show the versatility of our approach to
check properties that are handled using different approaches in the
literature (e.g., programming errors in the controller as in
\cite{DBLP:conf/pldi/BallBGIKSSV14}, safety policy violations as in
\cite{DBLP:conf/pldi/BallBGIKSSV14,DBLP:conf/fmcad/MajumdarTW14}, or
loop detection as in \cite{DBLP:conf/pldi/El-HassanyMBVV16}). And, on
the other hand, to show that we are able to handle networks larger than in related systems
\cite{DBLP:conf/fmcad/MajumdarTW14}, but without requiring
simplifications to the SDN models, nor extensions for DPOR reduction,
and in spite of using a non-distributed model checker. We should note
though that a precise comparison of figures is not possible due to the
differences described in Section~\ref{sec:dpor-reductions-sdn} and the
use of different implementations of controllers.

 Times are obtained on an Intel Core i7 at 3.4Ghz with 8GB of
RAM (Linux Kernel 3.2).  For each benchmark, we show in the second
column the number of switches, hosts and packets, \textbf{Execs}
corresponds to the number of different executions (i.e., branches in
the search tree),
\textbf{States} to the number of 
nodes in the search tree, and \textbf{Time} is the time taken by the
analysis in ms. Results are shown in Figure~\ref{fig:ex-results}.




\begin{figure}[p]
\centering

\begin{subfigure}{\textwidth}

\centering
 \begin{tabular}{rlrrr}\toprule 
 
 \multicolumn{1}{c}{\textbf{Name}} &
	      \multicolumn{1}{c}{\textbf{SxHxP}}
 & \multicolumn{1}{c}{\textbf{Execs}} 
 & \multicolumn{1}{c}{\textbf{States}} 
   &\multicolumn{1}{c}{\textbf{Time}}  \\ \toprule  
 
LB & 3x52x50 &  4 & 313 & 1305 \\
LB & 3x102x100 &  4 & 613 & 7301 \\
LB & 3x202x200 &  4 & 1213 & 38203 \\
LB & 3x302x300 &  4 & 1813 & 110220 \\
\hline \hline
LBB & 3x52x50 &  1 & 599 & 11117 \\
LBB & 3x77x75 &  1 & 899 & 31644 \\
LBB & 3x102x100 &  1 & 1199 & 68059 \\
LBB & 3x127x125 &  1 & 1499 & 127740 \\
\hline \hline

\toprule 
\end{tabular}
\captionsetup{skip=-.5ex}
\caption{Controller with load balancer.} 
\end{subfigure}
\bigskip

\begin{subfigure}{\textwidth}
\centering

 \begin{tabular}{rlrrr}\toprule 
 \multicolumn{1}{c}{\textbf{Name}} &
	      \multicolumn{1}{c}{\textbf{SxHxP}}
 & \multicolumn{1}{c}{\textbf{Execs}} 
 & \multicolumn{1}{c}{\textbf{States}} 
   &\multicolumn{1}{c}{\textbf{Time}}  \\ \toprule  
 
SSH & 2x2x100ssh &  1 & 407 & 83824 \\
SSH & 2x2x120oth &  1 & 490 & 146151 \\
SSH & 2x2x50each &  1 & 410 & 117245 \\
\hline \hline
SSH & 2x2x2cor &  179 & 1691 & 1340 \\
\hline \hline
SSHB & 2x2x2 &  6 & 120 & 104 \\
SSHB & 2x2x3 &  65 & 1419 & 2506 \\
SSHB & 3x3x4 & 421 & 10951 & 33470 \\
\hline \hline
\toprule
 \end{tabular}
\captionsetup{skip=-.5ex}
\caption{SSH controller.} 
\end{subfigure} 

\bigskip
 
\begin{subfigure}{\textwidth}
\centering

 \begin{tabular}{rlrrr}\toprule 
 \multicolumn{1}{c}{\textbf{Name}} &
	      \multicolumn{1}{c}{\textbf{SxHxP}}
 & \multicolumn{1}{c}{\textbf{Execs}} 
 & \multicolumn{1}{c}{\textbf{States}} 
   &\multicolumn{1}{c}{\textbf{Time}}  \\ \toprule  
 
LE & 3x3x2 &  3 & 71 & 42 \\
LE & 3x3x5 &  10 & 383 & 355 \\
LE & 6x3x2 &  5 & 217 & 272 \\
LE & 6x3x5 &  16 & 1040 & 2045 \\
LE & 9x2x2 &  10 & 787 & 4570 \\
LE & 15x2x2 &  16 & 2074 & 49274 \\
\hline \hline

\toprule
\end{tabular}	
\captionsetup{skip=-.5ex}
\caption{Network authentication with learning.}
\end{subfigure}

\bigskip

\begin{subfigure}{\textwidth}
\centering	

 \begin{tabular}{rlrrr}\toprule 
 \multicolumn{1}{c}{\textbf{Name}} &
	      \multicolumn{1}{c}{\textbf{SxHxP}}
 & \multicolumn{1}{c}{\textbf{Execs}} 
 & \multicolumn{1}{c}{\textbf{States}} 
   &\multicolumn{1}{c}{\textbf{Time}}  \\ \toprule  
 
MIb & 1x5x12 &  32 & 599 & 1029 \\
MIb & 1x5x14 &  64 & 1107 & 2730 \\
MIb & 1x5x16 &  748 & 9418 & 24870 \\
MIb & 1x8x20 &  2242 & 45539 & 153419 \\
\hline \hline
MI & 1x5x8 &  32 & 1004 & 865 \\
MI & 1x5x10 &  256 & 9436 & 9176 \\
MI & 1x5x12 &  960 & 17941 & 29675 \\
MI & 1x8x14 &  1727 & 55200 & 119908 \\
\hline \hline
\toprule
 \end{tabular}
\captionsetup{skip=-.5ex} 
 \caption{Firewall with migration.}
\end{subfigure}

\caption{Experimental results.}
\label{fig:ex-results}
\end{figure}

\paragraph{Controller with load balancer
  \cite{DBLP:conf/pldi/El-HassanyMBVV16} (LB/LBB)}\label{subsec:lb} 
 This corresponds to the controller of
\cite{DBLP:conf/pldi/El-HassanyMBVV16}, similar to our running
example.
\noindent It performs stateless load balancing among a set of replica
identified by a virtual IP (VIP) address. When receiving packets
destined to a VIP, the controller selects a particular host and
installs flow rules along the entire path.
For a buggy controller
without barriers (LB) and a network with 3 switches and 3 hosts, we
detect that there is a forwarding loop (i.e., that a packet reaches a
switch more than once) in 9ms after exploring 21 states.
For this, we have added to the
switches a field to store the packet identifiers that they have
already received, and when the same packet reaches it, it sends an
error message, which is observable from the final state.    
We are able
to scale this version up to 302 hosts and 300 packets. Once we
check the correct version with barriers (LBB), we are able to scale up
to 127 hosts and 125 packets. As it can be observed, for the largest
network, 1499 states are explored and in all cases we verify that the
traffic is balanced. The experiments in
\cite{DBLP:conf/pldi/El-HassanyMBVV16} do not specify the time to
detect the bug for this controller (they only mentioned that their
analysis finishes in less than 32s in the vast majority of
cases). Nevertheless, the underlying techniques to find the bugs are
unrelated (see Section~\ref{sec:concl-relat-work}), and thus time
comparison is not meaningful.

\paragraph{SSH controller \cite{DBLP:conf/fmcad/MajumdarTW14} (SSHE/SSHB)}
This case study is based on a controller that dynamically modifies the behaviors
\noindent of the switches as follows: 
it can update the switches with a rule
that states that no SSH packets are forwarded, and another that states
that all non-SSH packets are forwarded.
 We have two versions of the SSH controller.
The first three evaluations
correspond to an erroneous SSH controller that installs the rule to
forward packets and the rule to drop SSH packets with the same
priority, and thus the safety policy can be violated. As in \cite{DBLP:conf/fmcad/MajumdarTW14}, we evaluate
a network with 2 switches and 2 hosts.  As for packets,
we write \lst{100ssh}, \lst{120other}, and \lst{50each} to indicate that we
send 100 SSH packets, 120 non-SSH packets and 50 of each type. We
detect the error by checking in the switch if two contradictory drop
and forward packet actions are received for the same entry.  The
results that we obtain for 1 packet suggest higher performance of our approach: 
in \cite{DBLP:conf/fmcad/MajumdarTW14} they find the bug in 0.1s  and
we do it in 0.004s or 0.007s, depending on the type of packet. 
The last evaluation
\lst{2cor} corresponds to the correct SSH controller
for which we achieve a notable
improvement as we have now less tasks that match the same entry (as
priority is different). The row SSHB is a correct
implementation with barriers that reduces the number of executions for
2 packets notably because it guarantees that forward rules are
installed and thus switches will not send further requests. They
prove the correctness for SSHB-2-2 in 6.4  seconds  by  exploring  13
states, we explore 15 states (in 6ms) or 18 states (in 8 ms),
depending on the type of packet. Furthermore we are able to scale up to
3 hosts and 3 switches.

\secbeg \secbeg \secbeg

\paragraph{Network authentication with learning
  \cite{DBLP:conf/pldi/BallBGIKSSV14,DBLP:conf/fmcad/MajumdarTW14} (LE)}

This implements a composition of a learning switch with authentication
in \cite{DBLP:conf/pldi/BallBGIKSSV14}.
\noindent Also, \cite{DBLP:conf/fmcad/MajumdarTW14} evaluates a MAC learning
controller but using a different implementation.
LE implements a controller with barriers for which we can verify
flow-table consistency and that
the packet flows satisfy the intended policy. 
 We have considered configurations of 3x3, 6x3, 9x2 and 15x2.
When compared to \cite{DBLP:conf/fmcad/MajumdarTW14}, we
handle larger sizes of networks and for similar sizes, we explore less \textbf{States}
in less \textbf{Time}. We note that this might be due to the differences pointed out in
Section~\ref{sec:dpor-reductions-sdn} and different
implementations of the controller.


\paragraph{Firewall with migration\cite{DBLP:conf/pldi/BallBGIKSSV14} (MIb/MI)}
MI is the implementation of a firewall that supports migration of
trusted hosts. 
\noindent A host is trusted if it either sent/received (on some
switch) a message through/from port 1. 
Thus, when a trusted host
migrates to a new switch, the controller will remember it was trusted
before and will allow communication from either port.
For the same network 1x5 as \cite{DBLP:conf/pldi/BallBGIKSSV14}, we can scale
the number of packets up to 12 packets that actually modify the data
base for trusted hosts. We can keep on adding more packets if those do
not affect the shared data base. In MIb, we introduce the same bug in
the controller as \cite{DBLP:conf/pldi/BallBGIKSSV14}, which forgets
to check if trusted on events from port 2. We detect the
error by checking in the final state of the derivations that a packet
arrives to a host that is not in the trusted data base. The
scalability of MI and MIb are rather similar. However, we can handle
larger sizes of networks (1x8). 
Both
\cite{DBLP:conf/pldi/BallBGIKSSV14} and us find the bug in a negligible time.



\secbeg \secbeg \secbeg
\section{Conclusions}\label{sec:concl-relat-work}
\secbeg \subsecbeg 

We have proposed an actor-based framework to model and verify SDN
programs. A unique feature of our approach is that we can use
existing advanced verification algorithms without requiring any
specific extension to handle SDN features. This has allowed us to model and analyse several SDN scenarios: a controller with load balancer, an SSH controller, a learning switch with authentication, and a firewall with migration. Experiments have given evidence of the versatility and scalability of our approach.

We conclude with a review of related work in verification of software-defined networks and some directions for future work.

\subsection{Related work}


\paragraph{Static and Dynamic verification.}
The last years have
witnessed the development of many static and dynamic techniques for
verification that are closely related to our approach.
Static approaches have the main advantage that, when the
property can be proved, it is ensured for any possible execution,
while using dynamic analysis only guarantees the property for the
considered inputs. As a counterpart, in order to cover all possible
behaviors, static analysis needs to perform abstraction, which can give
a don't-know answer, and, possibly, false positives.
In \cite{DBLP:conf/pldi/BallBGIKSSV14}, the work on Horn-based
verification is lifted to the SDN programming paradigm, but excluding
barriers. Using this kind of verification, one can prove safety
invariants on the program. Our framework can additionally 
check liveness invariants (e.g., loop detection) by inspecting
 the traces computed by the model checker.
%
Static algebraic techniques are used in NetKAT \cite{DBLP:conf/popl/FosterKM0T15,AndersonFGJKSW14,BeckettGW16}, 
  to prove properties of SDN programs. 
NetKAT does not include primitives for concurrency, and has a significantly higher level of abstraction.
Therefore capturing features and scenarios we are interested in would be difficult.
 In \cite{DBLP:conf/sec/PascoalDFN17}, a particular type of attacks in
 the context of SDN networks has been modeled in Maude using the
 so-called hierarchically structured composite actor systems described
 in \cite{DBLP:conf/wadt/EckhardtMMW12}. This work does not provide a
 general model for SDN networks and, besides, barriers are not
 considered. On the other hand, it applies a statistical model
 checker, which requires to have a given scheduler for the
 messages. Such scheduler determines the exact order in which messages
 are handled while our framework captures all possible
 behaviours. Hence, both their aim and their SDN model are radically
 different from ours.

 Concerning dynamic techniques, our work is mostly related to the
 model checkers NICE and Kuai for SDN programs, which have been
 compared in detail in Section~\ref{sec:dpor-reductions-sdn}.
Our approach could be adapted 
to 
apply abstractions that bound the size of buffers
\cite{DBLP:conf/fmcad/MajumdarTW14} and to consider environment
messages \cite{DBLP:conf/fmcad/SethiNM13}.
 The approach of
\cite{DBLP:conf/pldi/El-HassanyMBVV16,DBLP:conf/nsdi/KazemianVM12} is based on analyzing dynamically given snapshots of
the network from
real executions.  Instead, we try to find programming
errors by inspecting only the SDN program and considering all possible
execution traces, thus enabling verification at system design time.

\paragraph{Data and Control-plane verification}
There is a substantial body of work on verification techniques for SDN focussing specifically on the data or the control plane. Data-plane approaches include: Anteater~\cite{MaiKACGK11}, which uses static analysis via SAT solving;
FlowChecker~\cite{Al-ShaerA10}, which applies symbolic model-checking to OpenFlow configurations;
VeriFlow~\cite{KhurshidZZCG13}, which provides an infrastructure to check data-plane properties in real-time. 
Control-plane approaches include:
Flowlog~\cite{NelsonFSK14}, a declarative language to program SDN controllers, which uses the Alloy model-checker to perform verification; \cite{NelsonFK15}, which uses differential analysis to discover bugs in different versions of the same controller program.

We stress that our approach targets both control and data-plane, and in particular it is capable of detecting bugs that arise from their interaction. 
Moreover, concurrency and barriers are not considered in the mentioned works.

\paragraph{Quantitative verification}
In~\cite{DBLP:conf/ifm/Galpin18}, SDN components are modelled via a quantitative process algebra. Their focus is on quantitative properties, e.g., latency and congestion. In particular, concurrency and barriers are not considered.

\paragraph{Network verification via actors}
Another actor-based verification framework is \emph{Rebeca}
(see~\cite{SirjaniJ11} for a survey). Rebeca supports a variety of
state-reduction techniques, and has been used to model and verify
wireless networks~\cite{YousefiGK17,DKhamespanahSMA18}.  Our approach
uses the ABS language and the SYCO tool. SYCO includes recent DPOR
techniques~\cite{DBLP:conf/cav/AlbertGIR18,DBLP:conf/cav/AlbertABGS17}
which, by exploiting specific features of SDNs, enabled us to better
scale and analyse larger networks.


\subsection{Future Work}

Although we did not explore it in this article, the encoding
we provide opens the door to apply a range of techniques other than
model checking. For instance, static analysis, runtime monitoring or
simulation of network behavior can be done now using the ABS toolsuite
\cite{abs-tools}. Other tools and methods for verification of
message-passing and concurrent-object systems could be also easily
adapted~\cite{DBLP:conf/icst/ChristakisGS13,Lauterburg:2010:BTS:1882291.1882349,DBLP:conf/popl/BouajjaniEEH15,DBLP:conf/popl/LiangF16}.
In addition, because the encoding is not very far from the original
flow tables, both model extraction from existing network code and code
generation from an actor model 
should be achievable with a small extension of the tool. 
This is left for future work.




~\newline
 \textbf{Acknowledgments} {This work was partially funded by the
   Spanish MECD Salvador de Madariaga Mobility Grants PRX17/00297 and
   PRX17/00303, the Spanish FPU Grant FPU15/04313, the Spanish MINECO
   projects TIN2015-69175-C4-2-R, TIN2015-69175-C4-3-R, the Spanish
   MCIU, AEI and FEDER (EU) through projects RTI2018-094403-B-C31 and
   RTI2018-094403-B-C33 and the CM project S2018/TCS-4314, the ERC
   starting grant Profoundnet (679127) and a Leverhulme Prize
   (PLP-2016-129).  }
       
 \bibliographystyle{elsarticle-num}
 \bibliography{biblio}

\begin{thebibliography}{10}
\expandafter\ifx\csname url\endcsname\relax
  \def\url#1{\texttt{#1}}\fi
\expandafter\ifx\csname urlprefix\endcsname\relax\def\urlprefix{URL }\fi
\expandafter\ifx\csname href\endcsname\relax
  \def\href#1#2{#2} \def\path#1{#1}\fi

\bibitem{actors}
G.~Agha, {A}ctors: {A} {M}odel of {C}oncurrent {C}omputation in {D}istributed
  {S}ystems, MIT Press, Cambridge, MA, 1986.

\bibitem{DBLP:conf/fmco/JohnsenHSSS10}
E.~B. Johnsen, R.~H{\"{a}}hnle, J.~Sch{\"{a}}fer, R.~Schlatte, M.~Steffen,
  {ABS:} {A} core language for abstract behavioral specification, in: FMCO,
  2010, pp. 142--164.

\bibitem{DBLP:conf/cc/AlbertGI16}
E.~Albert, M.~G{\'{o}}mez{-}Zamalloa, M.~Isabel, {SYCO:} a systematic testing
  tool for concurrent objects, in: {CC}, 2016, pp. 269--270.
\newblock \href {https://doi.org/10.1145/2892208.2892236}
  {\path{doi:10.1145/2892208.2892236}}.

\bibitem{AlbertGRSS18}
E.~Albert, M.~G{\'{o}}mez{-}Zamalloa, A.~Rubio, M.~Sammartino, A.~Silva,
  Sdn-actors: Modeling and verification of {SDN} programs, in: {FM}, 2018, pp.
  550--567.
\newblock \href {https://doi.org/10.1007/978-3-319-95582-7\_33}
  {\path{doi:10.1007/978-3-319-95582-7\_33}}.

\bibitem{DBLP:conf/cav/AlbertGIR18}
E.~Albert, M.~G{\'{o}}mez{-}Zamalloa, M.~Isabel, A.~Rubio, Constrained dynamic
  partial order reduction, in: {CAV}, 2018, pp. 392--410.
\newblock \href {https://doi.org/10.1007/978-3-319-96142-2\_24}
  {\path{doi:10.1007/978-3-319-96142-2\_24}}.

\bibitem{DBLP:conf/pldi/El-HassanyMBVV16}
A.~El{-}Hassany, J.~Miserez, P.~Bielik, L.~Vanbever, M.~T. Vechev, Sdnracer:
  concurrency analysis for software-defined networks, in: {POPL}, 2016, pp.
  402--415.
\newblock \href {https://doi.org/10.1145/2908080.2908124}
  {\path{doi:10.1145/2908080.2908124}}.

\bibitem{abs-tools}
{The ABS tool suite}, \url{http://abs-models.org}.

\bibitem{deboer07esop}
F.~S. de~Boer, D.~Clarke, E.~B. Johnsen, A {C}omplete {G}uide to the {F}uture,
  in: {ESOP}, Vol. 4421, 2007, pp. 316--330.

\bibitem{GuhaRF13}
A.~Guha, M.~Reitblatt, N.~Foster, Machine-verified network controllers, in:
  {PLDI}, 2013, pp. 483--494.
\newblock \href {https://doi.org/10.1145/2491956.2462178}
  {\path{doi:10.1145/2491956.2462178}}.

\bibitem{DBLP:conf/fase/SenA06}
K.~Sen, G.~Agha, Automated {S}ystematic {T}esting of {O}pen {D}istributed
  {P}rograms, in: {FASE}, 2006, pp. 339--356.

\bibitem{openflow}
Openflow switch specification, version 1.4.0 (October 2013).

\bibitem{DBLP:conf/nsdi/CaniniVPKR12}
M.~Canini, D.~Venzano, P.~Peres{\'{\i}}ni, D.~Kostic, J.~Rexford, A {NICE} way
  to test openflow applications, in: {NSDI}, 2012, pp. 127--140.

\bibitem{DBLP:conf/fmcad/MajumdarTW14}
R.~Majumdar, S.~D. Tetali, Z.~Wang, Kuai: {A} model checker for
  software-defined networks, in: {FMCAD}, 2014, pp. 163--170.
\newblock \href {https://doi.org/10.1109/FMCAD.2014.6987609}
  {\path{doi:10.1109/FMCAD.2014.6987609}}.

\bibitem{DBLP:conf/popl/FlanaganG05}
C.~Flanagan, P.~Godefroid, Dynamic partial-order reduction for model checking
  software, in: {POPL}, 2005, pp. 110--121.

\bibitem{DBLP:conf/forte/TasharofiKLLMA12-short}
S.~Tasharofi, R.~K. Karmani, S.~Lauterburg, A.~Legay, D.~Marinov, G.~Agha,
  Transdpor: {A} novel dynamic partial-order reduction technique for testing
  actor programs, in: {FMOODS/FORTE}, 2012, pp. 219--234.

\bibitem{DBLP:conf/cav/AlbertABGS17}
E.~Albert, P.~Arenas, M.~G. de~la Banda, M.~G{\'{o}}mez{-}Zamalloa, P.~J.
  Stuckey, Context-sensitive dynamic partial order reduction, in: {CAV}, Vol.
  10426, 2017, pp. 526--543.

\bibitem{DBLP:conf/pldi/BallBGIKSSV14}
T.~Ball, N.~Bj{\o}rner, A.~Gember, S.~Itzhaky, A.~Karbyshev, M.~Sagiv,
  M.~Schapira, A.~Valadarsky, Vericon: towards verifying controller programs in
  software-defined networks, in: {PLDI}, 2014, pp. 282--293.
\newblock \href {https://doi.org/10.1145/2594291.2594317}
  {\path{doi:10.1145/2594291.2594317}}.

\bibitem{DBLP:conf/popl/FosterKM0T15}
N.~Foster, D.~Kozen, M.~Milano, A.~Silva, L.~Thompson, A coalgebraic decision
  procedure for netkat, in: {POPL}, 2015, pp. 343--355.
\newblock \href {https://doi.org/10.1145/2676726.2677011}
  {\path{doi:10.1145/2676726.2677011}}.

\bibitem{AndersonFGJKSW14}
C.~J. Anderson, N.~Foster, A.~Guha, J.~Jeannin, D.~Kozen, C.~Schlesinger,
  D.~Walker, Netkat: semantic foundations for networks, in: {POPL}, 2014, pp.
  113--126.
\newblock \href {https://doi.org/10.1145/2535838.2535862}
  {\path{doi:10.1145/2535838.2535862}}.

\bibitem{BeckettGW16}
R.~Beckett, M.~Greenberg, D.~Walker, Temporal netkat, in: {PLDI}, 2016, pp.
  386--401.
\newblock \href {https://doi.org/10.1145/2908080.2908108}
  {\path{doi:10.1145/2908080.2908108}}.

\bibitem{DBLP:conf/sec/PascoalDFN17}
T.~A. Pascoal, Y.~G. Dantas, I.~E. Fonseca, V.~Nigam, Slow {TCAM} exhaustion
  ddos attack, in: SEC, 2017, pp. 17--31.

\bibitem{DBLP:conf/wadt/EckhardtMMW12}
J.~Eckhardt, T.~M{\"{u}}hlbauer, J.~Meseguer, M.~Wirsing, Statistical model
  checking for composite actor systems, in: WADT, 2012, pp. 143--160.

\bibitem{DBLP:conf/fmcad/SethiNM13}
D.~Sethi, S.~Narayana, S.~Malik, Abstractions for model checking {SDN}
  controllers, in: {FMCAD}, 2013, pp. 145--148.

\bibitem{DBLP:conf/nsdi/KazemianVM12}
P.~Kazemian, G.~Varghese, N.~McKeown, Header space analysis: Static checking
  for networks, in: {NSDI}, 2012, pp. 113--126.

\bibitem{MaiKACGK11}
H.~Mai, A.~Khurshid, R.~Agarwal, M.~Caesar, B.~Godfrey, S.~T. King, Debugging
  the data plane with anteater, in: {ACM} {SIGCOMM}, 2011, pp. 290--301.
\newblock \href {https://doi.org/10.1145/2018436.2018470}
  {\path{doi:10.1145/2018436.2018470}}.

\bibitem{Al-ShaerA10}
E.~Al{-}Shaer, S.~Al{-}Haj, Flowchecker: configuration analysis and
  verification of federated openflow infrastructures, in: {SafeConfig}, 2010,
  pp. 37--44.
\newblock \href {https://doi.org/10.1145/1866898.1866905}
  {\path{doi:10.1145/1866898.1866905}}.

\bibitem{KhurshidZZCG13}
A.~Khurshid, X.~Zou, W.~Zhou, M.~Caesar, P.~B. Godfrey, Veriflow: Verifying
  network-wide invariants in real time, in: {NSDI}, 2013, pp. 15--27.

\bibitem{NelsonFSK14}
T.~Nelson, A.~D. Ferguson, M.~J.~G. Scheer, S.~Krishnamurthi, Tierless
  programming and reasoning for software-defined networks, in: {NSDI}, 2014,
  pp. 519--531.

\bibitem{NelsonFK15}
T.~Nelson, A.~D. Ferguson, S.~Krishnamurthi, Static differential program
  analysis for software-defined networks, in: {FM}, 2015, pp. 395--413.
\newblock \href {https://doi.org/10.1007/978-3-319-19249-9\_25}
  {\path{doi:10.1007/978-3-319-19249-9\_25}}.

\bibitem{DBLP:conf/ifm/Galpin18}
V.~Galpin, Formal modelling of software defined networking, in: {IFM}, 2018,
  pp. 172--193.
\newblock \href {https://doi.org/10.1007/978-3-319-98938-9\_11}
  {\path{doi:10.1007/978-3-319-98938-9\_11}}.

\bibitem{SirjaniJ11}
M.~Sirjani, M.~M. Jaghoori, Ten years of analyzing actors: Rebeca experience,
  in: Formal Modeling: Actors, Open Systems, Biological Systems, 2011, pp.
  20--56.
\newblock \href {https://doi.org/10.1007/978-3-642-24933-4\_3}
  {\path{doi:10.1007/978-3-642-24933-4\_3}}.

\bibitem{YousefiGK17}
B.~Yousefi, F.~Ghassemi, R.~Khosravi, Modeling and efficient verification of
  wireless ad hoc networks, Formal Asp. Comput. 29~(6) (2017) 1051--1086.
\newblock \href {https://doi.org/10.1007/s00165-017-0429-z}
  {\path{doi:10.1007/s00165-017-0429-z}}.

\bibitem{DKhamespanahSMA18}
E.~Khamespanah, M.~Sirjani, K.~Mechitov, G.~Agha, Modeling and analyzing
  real-time wireless sensor and actuator networks using actors and model
  checking, {STTT} 20~(5) (2018) 547--561.
\newblock \href {https://doi.org/10.1007/s10009-017-0480-3}
  {\path{doi:10.1007/s10009-017-0480-3}}.

\bibitem{DBLP:conf/icst/ChristakisGS13}
M.~Christakis, A.~Gotovos, K.~F. Sagonas, {Systematic Testing for Detecting
  Concurrency Errors in Erlang Programs}, in: {ICST}, 2013, pp. 154--163.

\bibitem{Lauterburg:2010:BTS:1882291.1882349}
S.~Lauterburg, R.~K. Karmani, D.~Marinov, G.~Agha, Basset: a tool for
  systematic testing of actor programs, in: {SIGSOFT FSE}, 2010, pp. 363--364.
\newblock \href {https://doi.org/10.1145/1882291.1882349}
  {\path{doi:10.1145/1882291.1882349}}.

\bibitem{DBLP:conf/popl/BouajjaniEEH15}
A.~Bouajjani, M.~Emmi, C.~Enea, J.~Hamza, Tractable refinement checking for
  concurrent objects, in: POPL, 2015, pp. 651--662.
\newblock \href {https://doi.org/10.1145/2676726.2677002}
  {\path{doi:10.1145/2676726.2677002}}.

\bibitem{DBLP:conf/popl/LiangF16}
H.~Liang, X.~Feng, A program logic for concurrent objects under fair
  scheduling, in: POPL, 2016, pp. 385--399.
\newblock \href {https://doi.org/10.1145/2837614.2837635}
  {\path{doi:10.1145/2837614.2837635}}.

\end{thebibliography}

\newpage


\appendix

\section{Soundness Proofs}
\normalsize

\subsection{Proof of Theorem~\ref{th:modelling}}

Let $\alpha$ be the one-to-one function that pairs each element of $S$
with its unique actor in $S_a$ such that $S \equiv S_a$.

\newtheorem{assump}{Assumption}
\begin{assump}\label{assump:pol}
Given an SDN state $S=\langle H, Sw, C \rangle$ and an SDN-Actor state
$S_a$ such that $S \equiv S_a$, $C = c(top,cin)$ and
$a(c,tk,h,\queue) \in S_a$ with $\alpha(C) = a(c,tk,h,\queue)$, then $$applyPol(top,sid,o,ph) = applyPolicy(sid,o,ph)$$
\end{assump}

\begin{assump}
  $S^{ini} \equiv S_a^{ini}$, where $S_a^{ini}$ is the SDN-Actor state
  defined in Def.\ref{def:conf}, that is, the state after executing
  method \lst{init\_conf()} and all asynchronous calls
  to method \lst{sendIn} containing the packets to be delivered.
\end{assump}


\begin{definition}[succ($S$) (succ($S_a$))]
  Given an SDN(resp. SDN-Actor) state $S$ (resp. $S_a$), succ($S$)
  (resp. succ($S_a$)) denotes the set of final states of the
  executions in exec($S$) (resp. exec($S_a$)).
\end{definition}

\begin{lemma}Given an SDN state and an
SDN-Actor state $S_a$ such that $S \equiv S_a$ and $S \in
succ(S^{ini})$ and $S_a \in succ(S^{ini}_a)$,

\begin{enumerate} \item If $S \rightarrow S'$, then $\exists S_a'$ such that
$S_a \macrostep{} S_a'$ and $S' \equiv S_a'$.
\item If $S_a \macrostep{} S_a'$, then $\exists S'$ such that
$S \rightarrow S'$ and $S' \equiv S_a'$.
\end{enumerate}\label{lemma:equiv}
\end{lemma}

\begin{proof}
  Let us see reason about both points at the same time. We distinguish
  several cases depending on the semantics rule applied during the
  step $S \rightarrow S'$. Let $S$ and $S'$ be
  $\langle H, Sw, C \rangle$ and $\langle H', Sw', C' \rangle$,
  respectively.

\begin{enumerate}
\setlength{\itemsep}{3pt}

\item If rule (\textsc{si}) is applied,
      \begin{itemize}
        \item $H = H'' \cup \{h(id,sid,o,in\cup\{new(p)\})\}$ and $H' =
        H'' \cup \{h(id,sid,o,in)\}$
        \item $Sw {=} Sw'' \cup \{s(sid,\mathit{ft},b,in')\}$ and
        $Sw' {=} Sw'' \cup \{s(sid,\mathit{ft},b,in' \cup \{o{:}p\})\}$
        \item $C = C'$
      \end{itemize}

      Moreover, we know that $S \equiv S_a$, hence
      $\alpha(h(id,sid,o,in\cup\{new(p)\})) = a(hoid,\_,h,\queue)
      \in S_a$ and
      $\queue = \{\task(tid,\mbox{\lst{sendIn}},l,\_)\} \cup \queue'$. Thus, task
      $tid$ can be executed in state $S_a$ by actor $hoid$ and it will
      send a task \lst{shp} (where \lst{shp} stands for \lst{switchHandlePacket}) to $h[\mbox{\lst{s}}]$ such that
      $h[\mbox{\lst{s}}] = \alpha(s(sid,\mathit{ft},b,in')) = a(soid,\_,h2,\queue'') \in
      S_a$ (by the equivalence of $S$ and $S_a$).

      \begin{itemize}
        \item[a)] $S_a = S_a'' \cup \{a(hoid,\_,h,\queue),a(soid,\_,h2,\queue''),\}$
        \item[b)] $S_a' =
        S_a'' \cup \{a(hoid,\_,h,\queue'),a(soid,\_,h2,\queue''\cup\{\task(\_,\mbox{\lst{shp}},l2,\_)\})\}$,
        such that $l2[\mbox{\lst{p}}] = l[\mbox{\lst{p}}]$ and $l2[\mbox{\lst{o}}] = h[\mbox{\lst{o}}]$.
      \end{itemize}

      By such equivalence we know that (1)
      $\langle H'',Sw'',C \rangle \equiv S_a''$, (2)
      $(in,\emptyset)\equiv (\queue',\emptyset)$ and by a) and b)
      $(in' \cup \{o:p\},\emptyset) \equiv (\queue'' \cup
      \{\task(\_,\mbox{\lst{shp}},l2,\_)\},\emptyset)$ Consequently,
      $S' \equiv S_a'$.

    \item If rule (\textsc{hhp}) is applied,
      
      \begin{itemize}
        \item $H = H'' \cup \{h(id,sid,o,in\cup\{p\})\}$ and $H' =
        H'' \cup \{h(id,sid,o,in)\}$
        \item $Sw = Sw'$ and $C = C'$
        \end{itemize}
        
      We also know $S \equiv S_a$, hence
      $\alpha(h(id,sid,o,in\cup\{p\})) = a(hoid,\_,h,\queue) \in S_a$
      and $\queue = \{\task(tid,\mbox{\lst{hhp}},l,\_)\} \cup
      \queue'$ (where \lst{hhp} stands for \lst{hostHandlePacket}). Therefore task $tid$ can be executed in state $S_a$ by
      $hoid$ and the packet is removed from the buffer.

      \begin{itemize}
      \item $S_a = S_a'' \cup \{a(hoid,\_,h,\queue)\}$
      \item $S_a' = S_a'' \cup \{a(hoid,\_,h,\queue')\}$.
      \end{itemize}

      Furthermore, $p = l[p]$.  By the equivalence between $S_a$ and
      $S$, we know that
      $(in \cup \{p\}, \emptyset) \equiv (\queue, \emptyset)$ and
      then, (1) $(in,\emptyset) \equiv (\queue',\emptyset)$ and also,
      (2) $\langle H'',Sw,C \rangle \equiv S_a''$. Consequently,
      $S' \equiv S_a'$.

\item If rule (\textsc{shp$_1$}) is applied,

  \begin{itemize}
  \item $H = H'' \cup \{h(id,sid,o,in'\}$ and
    $H'=H'' \cup \{h(id,sid,o,in'\cup \{p\})\}$
  \item $Sw {=} Sw'' \cup \{s(sid,\mathit{ft},b,in \cup\{o2:p\})\}$ and
    $Sw'{=} Sw'' \cup \{s(sid,\mathit{ft},b,in)\}$
  \item $C = C'$ and
    $\langle send(id)\rangle = lookup(\mathit{ft},\langle header(p),o
    \rangle)$.
  \end{itemize}

  Moreover, we know $S \equiv S_a$, hence
  $\alpha(s(sid,\mathit{ft},b,in)) = a(soid,\_,h,\queue) \in S_a$ and
  $\queue = \{\task(tid,\mbox{\lst{shp}},l,\_)\} \cup \queue'$.
  Therefore, task $tid$ can be executed and by the equivalence, we
  know that $\mathit{ft} = h[\mbox{\lst{flowT}}]$, thus the action returned by $\mbox{\lst{flowT}}$ is
  the same that the one returned by $\mathit{ft}$, that is
  $send(id)$. Therefore, the check in line $21$ does not succeed, but
  the check in line $22$ does, and, consequently, it spawns a task
  \lst{hhp} to
  $\alpha(h(id,sid,o,in')) = a(hoid,\_,h2,\queue'') \in S_a$. As a
  result
  
  \begin{itemize}
  \item
    $S_a = S_a'' \cup \{a(hoid,\_,h2,\queue''),a(soid,\_,h,\queue)\}$
  \item
    $S_a' = S_a'' \cup
    \{a(hoid,\_,h2,\queue''\cup\{\task(\_,\mbox{\lst{hhp}},l2,\_)\}),a(soid,\_,h,\queue')\}$
  \end{itemize}
        
  where $l[\mbox{\lst{p}}]=l2[\mbox{\lst{p}}]$. We also know by $S \equiv S_a$, that (1)
  $\langle H'',Sw'',C \rangle \equiv S_a''$, (2)
  $(in \cup \{o2:p\},b) \equiv (\queue,h[\text{\mbox{\lst{buffer}}}])$. Since
  $p=l2[p]$, then
  $(in' \cup \{p\},b) \equiv
  (\queue''\cup\{\task(\_,\mbox{\lst{hhp}},l2,\_)\},h[\text{\mbox{\lst{buffer}}}])$.
  Consequently, we have that $S' \equiv S_a'$.

\item If rule (\textsc{shp$_2$}) is applied,
        
  \begin{itemize}
  \item
    $Sw = Sw'' \cup \{s(sid,\mathit{ft},b,in
    \cup\{o2:p\}),s(sid',\mathit{ft}',b',in')\}$
  \item
    $Sw' = Sw'' \cup \{s(sid,\mathit{ft},b,in),s(sid',\mathit{ft}',b',in'\cup\{o:p\})\}$
  \item $H = H', ~C = C'$ and
    $send(sid',o) = lookup(\mathit{ft},\langle header(p),o2 \rangle)$.
  \end{itemize}

  Moreover, we know that $S \equiv S_a$, so
  $\alpha(s(sid,\mathit{ft},b,in)) = a(soid,\_,h,\queue) \in S_a$ and
  $\queue = \{\task(tid,\mbox{\lst{shp}},l,\_)\} \cup \queue'$.  Hence,
  task $tid$ can be executed, and by the equivalence, we know that
  $\mathit{ft} = h[\mbox{\lst{flowT}}]$, thus the action returned by $\mbox{\lst{flowT}}$ is the same
  that the one returned by $\mathit{ft}$, that is $ send(soid',o)$. Therefore,
  the check in line $21$ succeeds and, consequently, it spawns a task
  \lst{shp} to
  $\alpha(s(sid',\mathit{ft}',b',in')) = a(soid',\_,h2,\queue'') \in S_a$. As a
  result
  
  \begin{itemize}
  \item
    $S_a = S_a'' \cup \{a(soid',\_,h2,\queue''),a(soid,\_,h,\queue)\}$
  \item
    $S_a' = S_a'' \cup
    \{s(soid',\_,h2,\queue''\cup\{\task(\_,\mbox{\lst{shp}},l2,\_)\}),a(soid,\_,h,\queue')\}$
  \end{itemize}

  where $l[p]=l2[p]$. We also know by $S \equiv S_a$, that (1)
  $\langle H,Sw'',C \rangle \equiv S_a''$, (2)
  $(in \cup \{o2:p\},b) \equiv (\queue,h[\text{\mbox{\lst{buffer}}}])$. Finally, we
  also know that
  $(in' \cup \{o:p\},b') \equiv
  (\queue''\cup\{\task(\_,\mbox{\lst{shp}},l2,\_)\},h2[\text{\mbox{\lst{buffer}}}])$
  because $o = l2[\mbox{\lst{o}}], p=l2[\mbox{\lst{p}}]$. Consequently, we have that
  $S' \equiv S_a'$.

\item If rule \textsc{shp$_3$} is applied,
  
  \begin{itemize}
  \item $Sw = Sw'' \cup \{s(sid,\mathit{ft},b,in\cup\{o:p\})\}$ and
    $Sw = Sw'' \cup \{s(sid,\mathit{ft},b\cup\{o:p\},in)\}$
  \item $C = c(top,in')$ and
    $C' = c(top,in'\cup\{\textsf{pktIn(sid,o,id(p),header(p))}\})$
  \item $H = H'$ and $\bot = lookup(\mathit{ft},\langle header(p),o\rangle)$.
  \end{itemize}

  Moreover, we know that $S \equiv S_a$, so
  $\alpha(s(sid,\mathit{ft},b,in)) = a(soid,\_,h,\queue) \in S_a$ and
  $\queue = \{\task(tid,\mbox{\lst{shp}},l,\_)\} \cup \queue'$.
  Hence, task $tid$ can be executed and by the equivalence, we know
  that $\mathit{ft} = h[\mbox{\lst{flowT}}]$, so the action returned by $\mbox{\lst{flowT}}$ is the same
  that the one returned by $\mathit{ft}$, that is $ \bot$. Therefore, the
  checks in line $21$ and $22$ do not succeed and, consequently, it
  (1) spawns a task \lst{chm} (where \lst{chm} stands for\lst{controlHandleMessage}) to
  $h[ctrl] = \alpha(c(top,in')) = a(coid,\_,h2,\queue'') \in S_a$,
  and, (2) it stores the packet and the port $o:p$ in
  $h[\text{\mbox{\lst{buffer}}}]$.  As a result

  \begin{itemize}
  \item
    $S_a = S_a'' \cup \{a(soid,\_,h,\queue),a(coid,\_,h2,\queue'')\}$.
  \item
    $S_a' = S_a'' \cup \{a(soid,\_,h',\queue'),a(coid,\_,h2,\queue''
    \cup \{\task(\_,\mbox{\lst{chm}},$ $l2,\_)\})\}$ and $h' := h$
    but $h'[\text{\mbox{\lst{buffer}}}] := h[\text{\mbox{\lst{buffer}}}]\cup\{(p,o)\}$.
  \end{itemize}
  Furthermore, we know that (1)
  $\langle H,Sw'',C \rangle \equiv
  (S_a''\cup\{a(coid,\_,h2,\queue'')\})$ and (2)
  $(in,b\cup\{o:p\}) \equiv (\queue',h[\text{\mbox{\lst{buffer}}}]\cup\{(p,o)\})
  $. Moreover, we have $(in',\emptyset) \equiv (\queue''\cup \{task$
  $(\_,\mbox{\lst{chm}},l2,\_\},\emptyset)$ since
  $soid = l2[\mbox{\lst{sid}}],~ p = l[\mbox{\lst{p}}]$ and $o = l[\mbox{\lst{o}}]$.  Consequently, we have
  $S' \equiv S_a'$.

\item If rule (\textsc{so$_1$}) is applied,
  \begin{itemize}
  \item $H = H'' \cup \{h(id,sid,o,in'\}$ and
    $H'=H'' \cup \{h(id,sid,o,in'\cup \{o:p\})\}$
  \item
    $Sw = Sw'' \cup \{s(sid,\mathit{ft},b\cup\{o2:p\},in \cup\{\textsf{pktOut(ph)}\})\}$
    and $Sw' = Sw'' \cup \{s(sid,\mathit{ft},b,in)\}$
  \item $C = C'$, $ph= header(p)$ and
    $send(id) = lookup(\mathit{ft},\langle header(p), o\rangle)$.
  \end{itemize}

  We also know $S \equiv S_a$, so
  $\alpha(s(sid,\mathit{ft},b\cup\{o2:p\},in\cup\{\textsf{pktOut(ph)}\})) =
  a(soid,\_,h,\queue) \in S_a$ and
  $\queue = \{\task(tid,\mbox{\lst{sendOut}},l,\_)\} \cup \queue'$.  Hence, task
  $tid$ can be executed and by the equivalence, we know that
  $\mathit{ft} = h[\mbox{\lst{flowT}}]$, so the action returned by $\mbox{\lst{flowT}}$ is the same that
  the one returned by $\mathit{ft}$, that is $send(soid,o)$.  Therefore, the
  check in line $28$ does not succeed, but the check in line $29$
  does, and, consequently, it spawns a task \lst{hhp} to
  $\alpha(h(id,sid,o,in')) = a(hoid,\_,h2,\queue'') \in S_a$. As a
  result

  \begin{itemize}
  \item
    $S_a = S_a'' \cup \{a(hoid,\_,h2,\queue''),a(soid,\_,h,\queue)\}$
  \item
    $S_a' = S_a'' \cup
    \{a(hoid,\_,h2,\queue''\cup\{\task(\_,\mbox{\lst{hhp}},l2,\_)\}),a(soid,\_,h',\queue')\}$
  \end{itemize}

  where $h' := h$ but
  $h'[\text{\mbox{\lst{buffer}}}] := take(h[\text{\mbox{\lst{buffer}}}],l[\mbox{\lst{pi}}])$. We also know by
  $S \equiv S_a$, that (1) $\langle H'',Sw'',C \rangle \equiv S_a''$,
  (2) $(in \cup \{o2:p\},b) \equiv
  (\queue,h[\text{\mbox{\lst{buffer}}}])$. Finally, we also know that
  $(in' \cup \{o:p\},b) \equiv
  (\queue''\cup\{\task(\_,\mbox{\lst{hhp}},l2,\_)\},h[\text{\mbox{\lst{buffer}}}])$
  since $(p,o) = take(h[\text{\mbox{\lst{buffer}}}],l[\mbox{\lst{pi}}]) = (l2[\mbox{\lst{p}}],o)$.
  Consequently, we have that $S' \equiv S_a'$.

\item If rule (\textsc{so$_2$}) is applied,
  \begin{itemize}
  \item
    $Sw = Sw'' \cup \{s(sid,\mathit{ft},b\cup\{o2:p\},in
    \cup\{\textsf{pktOut(ph)}\}),s(sid',\mathit{ft}',b',in')\}$
  \item
    $Sw' = Sw'' \cup \{s(sid,\mathit{ft},b,in),s(sid',\mathit{ft}',b',in'\cup\{o:p\})\}$
  \item $H = H', ~C = C'$ and
    $send(sid',o) = lookup(\mathit{ft},\langle header(p),o2 \rangle)$.
  \end{itemize}
  
  Moreover, we know $S \equiv S_a$, so
  $\alpha(s(sid,\mathit{ft},b\cup\{o2:h\},in\cup\{\textsf{pktOut(ph)}\})) =
  a(soid,\_,h,\queue) \in S_a$ and
  $\queue = \{\task(tid,\mbox{\lst{sendOut}},l,\_)\} \cup \queue'$.  Hence, task
  $tid$ can be executed and, by the equivalence, we know that
  $\mathit{ft} = h[\mbox{\lst{flowT}}]$, so the action returned by $\mbox{\lst{flowT}}$ is the same that
  the one returned by $\mathit{ft}$, that is $send(soid',o)$. Therefore, the
  check in line $28$ succeeds, and consequently, it spawns a task
  \lst{shp} to
  $\alpha(s(sid',\mathit{ft}',b',in')) = a(soid',\_,h2,\queue'') \in S_a$. As a
  result
  
  \begin{itemize}
  \item
    $S_a = S_a'' \cup \{a(soid',\_,h2,\queue''),a(soid,\_,h,\queue)\}$
  \item
    $S_a' = S_a'' \cup
    \{s(soid',\_,h2,\queue''\cup\{\task(\_,\mbox{\lst{shp}},l2,\_)\}),a(soid,\_,h,\queue')\}$
  \end{itemize}

  where $h' := h$ but
  $h'[\text{\mbox{\lst{buffer}}}] := take(h[\text{\mbox{\lst{buffer}}}],l[\mbox{\lst{pi}}])$. We also know by
  $S \equiv S_a$, that (1) $\langle H,Sw'',C \rangle \equiv S_a''$,
  (2) $(in \cup \{o2:p\},b) \equiv
  (\queue,h[\text{\mbox{\lst{buffer}}}])$. Finally, we also know that
  $(in' \cup \{o:p\},b) \equiv
  (\queue''\cup\{\task(\_,\mbox{\lst{shp}},l2,\_)\},h[\text{\mbox{\lst{buffer}}}])$
  since $(p,o) = take(h[\text{\mbox{\lst{buffer}}}],l[\mbox{\lst{pi}}]) = (l2[\mbox{\lst{p}}],l2[\mbox{\lst{o}}])$.
  Consequently, we have that $S' \equiv S_a'$.

\item If rule (\textsc{so$_3$}) is applied,
  \begin{itemize}
  \item
    $Sw = Sw'' \cup \{s(sid,\mathit{ft},b\cup\{o:p\},in \cup\{\textsf{pktOut(ph)}\})\}$
  \item $Sw'= Sw'' \cup \{s(sid,\mathit{ft},b,in)\}$
  \item $H{=} H', ~C {=} C'$, and $ph {=} header(p)$ and
    $\bot {=} lookup(\mathit{ft},\langle header(p),o \rangle)$.
  \end{itemize}
  Moreover, we know that $S \equiv S_a$, so
  $\alpha(s(sid,\mathit{ft},b\cup\{o{:}p\},in\cup\{\textsf{pktOut(ph)}\}))$
  $=
  a(soid,\_,h,\queue) \in S_a$ and
  $\queue = \{\task(tid,\mbox{\lst{sendOut}},l,\_)\} \cup \queue'$.  Hence, task
  $tid$ can be executed, and by the equivalence, we know that
  $\mathit{ft} = h[\mbox{\lst{flowT}}]$, thus the action returned by $\mbox{\lst{flowT}}$ is the same
  that the one returned by $\mathit{ft}$, that is $ \bot$. Therefore, the
  checks in line $28$ and $29$ do not succeed, and consequently, it
  drops the packet without spawning any other task. As a result

  \begin{itemize}
  \item $S_a = S_a'' \cup \{a(soid,\_,h,\queue)\}$
  \item $S_a' = S_a'' \cup \{a(soid,\_,h',\queue')\}$
  \end{itemize}
  
  where $h' := h$ but
  $h'[\text{\mbox{\lst{buffer}}}] := take(h[\text{\mbox{\lst{buffer}}}],l[\mbox{\lst{pi}}])$. We also know by
  $S \equiv S_a$, that (1) $\langle H,Sw'',C \rangle \equiv S_a''$,
  (2) $(in,b) \equiv (\queue',h'[\text{\mbox{\lst{buffer}}}])$.  Consequently, we
  have that $S' \equiv S_a'$.

\item If rule (\textsc{shm}) is applied,

  \begin{itemize}
  \item
    $Sw = Sw'' \cup \{s(sid,\mathit{ft},b,in \cup\{\textsf{modState}(\langle \textsf{ph, o}
    \rangle\mapsto \textsf{a})\})\}$
  \item
    $Sw' = Sw'' \cup \{s(sid,put(\mathit{ft},\langle ph, o \rangle,a),b,in)\}$
  \item $H = H', ~C = C'$.
  \end{itemize}

  Moreover, we know that $S \equiv S_a$, so
  $\alpha(s(sid,\mathit{ft},b,in\cup\{\textsf{modState}(\langle \textsf{ph,o} \rangle \mapsto
  \textsf{a})\}))= a(soid,\_,h,\queue) \in S_a$ and
  $\queue = \{\task(tid,\mbox{\lst{switchHandleMessage}},l,\_)\} \cup \queue'$.
  Therefore, task $tid$ can be executed, and by the equivalence, we
  know that $\mathit{ft} = h[\mbox{\lst{flowT}}]$, hence
  $put(\mathit{ft},m,a) = put(h[\mbox{\lst{flowT}}],\langle ph,o \rangle,a)$, and
  $m = \langle ph,o \rangle $ because of Assumption \ref{assump:pol},
  that is, \textit{applyPol} and \textit{applyPolicy} behaves
  similarly for $\alpha(s)$ and $s$, $\forall s \in Sw$ and
  $\alpha(s) \in S$.

  \begin{itemize}
  \item $S_a = S_a'' \cup \{a(soid,\_,h,\queue)\}$
  \item $S_a' = S_a'' \cup \{a(soid,\_,h',\queue')\}$
  \end{itemize}

  where $h' := h$ but $h'[\mbox{\lst{flowT}}] := put(h[\mbox{\lst{flowT}}],m,a)$. We also know
  by $S \equiv S_a$, that (1) $\langle H,Sw'',C \rangle \equiv S_a''$,
  (2) $(in,b) \equiv (\queue',h[\text{\mbox{\lst{buffer}}}])$.  Consequently, we
  have that $S' \equiv S_a'$.

\item If rule (\textsc{chm}) is applied,

  \begin{itemize}
  \item $Sw = Sw'' \cup \{s(sid,\mathit{ft},b,in)\}$
  \item
    $Sw' = Sw_{ms}'' \cup \{s(sid,\mathit{ft},b,in\cup ms_{sid} \cup
    \{\textsf{pktOut(ph)}\}\}$,
  \item $H = H'$, $C = c(top,cin\cup\{\textsf{pktIn(sid,o,pid,ph)}\})$ and
    $C' = c(top,cin)$
  \end{itemize}

  where $ms = applyPol(top,sid,o,ph)$ and
  $ms_{id} = \{m | \langle id, m \rangle \in ms\}$.  Moreover, we know
  that $S \equiv S_a$, so
  $\alpha(c(top,cin\cup\{\textsf{pktIn(sid,o,pid,ph)}\})) = a(coid,\_,h,\queue)
  \in S_a$ and
  $\queue = \{\task(tid,\mbox{\lst{chm}},l1,\_)\} \cup \queue'$ (where
  \lst{chm} stands \mbox{for} \lst{chm})
  such that $soid = l1[\mbox{\lst{sid}}],~o=l1[\mbox{\lst{o}}],~pid = l1[\mbox{\lst{p}}]$ and $ph = l1[\mbox{\lst{h}}]$.
  Therefore, task $tid$ can be executed, and by the equivalence of $S$
  and $S_a$ and Assumption \ref{assump:pol}, we know that the list $l$
  is equivalent to $ms$, in the sense that it contains exactly the
  same switches and the same actions. Hence, in line $41$, actor
  $coid$ spawns tasks \lst{shm} to every switch in the
  list $l$ (where \lst{shm} stands
  for \lst{switchHandleMessage}) and finally, it spawns a a task \lst{sendOut} to actor
  $soid$ in line $43$. Let us see the equivalence between $S_a'$ and
  $S'$.
  
  \begin{itemize}
  \item $\forall s(sid',\mathit{ft}',b',in') \in Sw''$ such that
    $ms_{sid'} = \emptyset$, then $s(sid,\mathit{ft}',b',in') \in
    Sw_{ms}''$. Furthermore, if $ms_{sid'} = \emptyset$, then
    $sid'\not\in l$, hence $sid'$ will not receive any
    message. Therefore,
    $\alpha(s(sid',\mathit{ft}',b',in')) \in S_a \cap S_a'$.
  \item $\forall s(sid',\mathit{ft}',b',in') \in Sw''$ such that
    $ms_{sid'} \neq \emptyset$,then
    $s(sid,\mathit{ft}',b',in' \cup ms_{sid'}) \in Sw_{ms}''$. Then, $coid$
    will spawn as many \lst{shm} tasks as messages in
    $\{(m,a) | (soid',m,a) \in l\}$ (and in $ms_{soid'}$).  By $S_a \equiv S$, we know
    $\alpha(s(sid,\mathit{ft}',b',in')) = a(soid',\_,h2,\queue'') \in S_a$ and
    $(in',b') \equiv (\queue'',h2[\text{\mbox{\lst{buffer}}}])$. Furthermore,
    $a(soid',\_,h2,\queue''\cup tks_{l,soid'}) \in S_a'$, where
    $tks_{l,soid'}{:=}\{ \task(\_,\mbox{\lst{shm}},l',\_)$ $ |
    (soid',m,a) {\in} l, l'[\mbox{\lst{m}}]{:=}m,~l'[\mbox{\lst{a}}]{:=}a\}$ which is equivalent
    to the information contained in $ms_{sid'}$. Then,
    $(in'\cup ms_{sid'},b) \equiv (\queue'' \cup
    tks_{l,soid'},h2[\text{\mbox{\lst{buffer}}}])$.
  \item Regarding the switch
    $s(sid,\mathit{ft},b,in)$, by the equivalence of $S$ and
    $S_a$, we know that $\alpha(s(sid,\mathit{ft},b,in)) =
    a(soid,\_,h1,\queue_{soid}) \in S_a$ and since $soid =
    l[\mbox{\lst{sid}}]$, actor
    $coid$ spawns a task \lst{sendOut}, and as many
    \lst{shm} tasks as messages in $\{(m,a) |
    (soid,m,a) \in l\}$, and then $a(soid,\_,h1,\queue_{soid}\cup
    tks_{l,soid} \cup \{\task(\_,\mbox{\lst{sendOut}},l',\_)\}) \in
    S_a'$. Again, by the equivalence of $S$ and
    $S_a$, we know that $(in,b) \equiv
    (\queue_{soid},h1[\text{\mbox{\lst{buffer}}}])$ and, since
    $tks_{l,soid}$ is the equivalent information contained in
    $ms_{sid}$, then we get that $(in\cup ms_{sid},b) \equiv
    (\queue_{soid} \cup
    tks_{l,soid},h1[\text{\mbox{\lst{buffer}}}])$. Finally,
    $\textsf{pktOut(ph)}$ contains the equivalent information to
    $t_{out}{:=}\task(\_,\mbox{{\lst{sendOut}}},l',\_)$, thus we get $(in'\cup
    (ms_{sid'} \cup \{\textsf{pktOut(ph)}\},b) \equiv (\queue'' \cup
    tks_{l,soid'} \cup \{t_{out}\},h[\text{\mbox{\lst{buffer}}}])$.
  \item Regarding the controller $C$, we know that $(cin \cup
    \{\textsf{pktIn(sid,o,pid,ph)}\},\emptyset) \\\equiv (\queue \cup
    \{\task(tkid,\mbox{\lst{chm}},l1,\_)\},\emptyset)$. Furthermore,
    by Assumption \ref{assump:pol}, we know that
    $related(top,\{h[\textsf{srefs},\textsf{href},\textsf{ntw}]\})$. As a consequence
    $(cin,\emptyset) \\\equiv (\queue,\emptyset)$.
  \end{itemize}
  All in all, we conclude that $S' \equiv S_a'$.
\end{enumerate}

Let us notice here that even though we have distinguished the
different cases depending on the semantics rule for SDN networks, the
previous reasoning also includes each possible execution of a task in
the SDN-Actor model. Hence, each possible execution of a task
corresponds exactly with one of the semantics rule for SDN networks.
\end{proof}

\newenvironment{Reptheorem}[1]{%
  \renewcommand\thetheorem{#1}
  \theorem
}{\endtheorem}

\begin{Reptheorem}{\ref{th:modelling}}
Let $S^{ini}$ and $S_a^{ini}$ be an SDN state and an SDN-Actor state,
respectively.

\begin{enumerate}
\item For every execution $S^{ini} \rightarrow S^1 \rightarrow... \rightarrow
S^n \in exec(S^{ini}), \exists
S^{ini}_a \macrostep{} S_a^1 \macrostep{} ... \macrostep{} S_a^n \in
exec(S_{a}^{ini})$ such that $S^n \equiv S_a^n$.
\item For every execution $S_a^{ini} \macrostep{} S_a^1 \macrostep{}... \macrostep{}
S_a^n \in exec(S_a^{ini}), \exists
S^{ini} \rightarrow S^1 \rightarrow ... \rightarrow S^n \in
exec(S^{ini})$ such that $S^n \equiv S_a^n$.
\end{enumerate}
\end{Reptheorem}

\begin{proof}
Let us prove both cases by induction on the length $n$ of the
execution.

\begin{itemize}
\item If $n=0$, it is straightforward to see that $S^0 = S^{ini} \equiv
S_a^{ini} = S_a^0 $.
\item Let us suppose that both cases are true for $n$ and let us prove
  them for $n+1$
  
\begin{enumerate}
\item We need to prove that for every execution
  $S^{ini} \rightarrow S^1 \rightarrow... \rightarrow S^{n+1} \in
  exec(S^{ini}), $
  $\exists S^{ini}_a \macrostep{} S_a^1 \macrostep{} ... \macrostep{}
  S_a^{n} \macrostep{} S_a^{n+1} \in exec(S_{a}^{ini})$ such that
  $S^{n+1} \equiv S_a^{n+1}$. Applying the induction hypothesis we
  know that
  $\exists S^{ini}_a \macrostep{} S_a^1 \macrostep{} ... \macrostep{}
  S_a^{n} \in exec(S_{a}^{ini})$ such that $S^{n} \equiv
  S_a^{n}$. Therefore, now we have $S^n \rightarrow S^{n+1}$, and
  $S^n \equiv S_a^n$, hence, applying Lemma \ref{lemma:equiv}.1 we get
  that $\exists S_a^{n+1}$ such that $S_a^n \macrostep{} S_a^{n+1}$
  and $S^{n+1} \equiv S_a^{n+1}$.
\item We need to prove that for every execution
  $S_a^{ini} \macrostep{} S_a^1 \macrostep{}... \macrostep{} S_a^{n+1}
  \in exec(S_a^{ini}),$
  $ \exists S^{ini} \rightarrow S^1 \rightarrow ... \rightarrow S^{n}
  \rightarrow S^{n+1} \in exec(S^{ini})$ such that
  $S^{n+1} \equiv S_a^{n+1}$. Applying the induction hypothesis we
  know that
  $\exists S^{ini} \rightarrow S^1 \rightarrow ... \rightarrow S^{n}
  \in exec(S^{ini})$ such that $S^{n} \equiv S_a^{n}$. Therefore, now
  we have $S_a^n \macrostep{} S_a^{n+1}$, and $S^n \equiv S_a^n$,
  hence, applying Lemma \ref{lemma:equiv}.2 we get 
  $\exists S^{n+1}$ such that $S^n \rightarrow S^{n+1}$ and
  $S^{n+1} \equiv S_a^{n+1}$.

\end{enumerate}
\end{itemize}
\end{proof}

\end{document}